\input harvmac
\input psfig
\newcount\figno
\figno=0
\def\fig#1#2#3{
\par\begingroup\parindent=0pt\leftskip=1cm\rightskip=1cm\parindent=0pt
\global\advance\figno by 1
\midinsert
\epsfxsize=#3
\centerline{\epsfbox{#2}}
\vskip 12pt
{\bf Fig. \the\figno:} #1\par
\endinsert\endgroup\par
}
\def\figlabel#1{\xdef#1{\the\figno}}
\def\encadremath#1{\vbox{\hrule\hbox{\vrule\kern8pt\vbox{\kern8pt
\hbox{$\displaystyle #1$}\kern8pt}
\kern8pt\vrule}\hrule}}

\overfullrule=0pt

%
\def\underarrow#1{\vbox{\ialign{##\crcr$\hfil\displaystyle
 {#1}\hfil$\crcr\noalign{\kern1pt\nointerlineskip}$\longrightarrow$\crcr}}}
%
\def\tilde{\widetilde}
\def\bar{\overline}

\def\inbar{\vrule height1.5ex width.4pt depth0pt}
\def\IC{\relax\hbox{\kern.25em$\inbar\kern-.3em{\rm C}$}}
\def\IR{\relax\hbox{\kern.25em$\inbar\kern-.3em{\rm R}$}}
\def\IZ{\relax\ifmmode\hbox{Z\kern-.4em Z}\else{Z\kern-.4em Z}\fi}

\font\ninerm=cmr8
\font\zfont = cmss10 

\def\bigone{\hbox{1\kern -.23em {\rm l}}}
\def\ZZ{\hbox{\zfont Z\kern-.4emZ}}

\def\RR{{\cal R}}
\def\NN{{\cal N}}

\def\ov{\overline}

\def\drawbox#1#2{\hrule height#2pt
        \hbox{\vrule width#2pt height#1pt \kern#1pt
              \vrule width#2pt}
              \hrule height#2pt}

\def\Fund#1#2{\vcenter{\vbox{\drawbox{#1}{#2}}}}
\def\Asym#1#2{\vcenter{\vbox{\drawbox{#1}{#2}
              \kern-#2pt       
              \drawbox{#1}{#2}}}}

\def\fund{\Fund{6.5}{0.4}}

\batchmode
  \font\bbbfont=msbm10
\errorstopmode
\newif\ifamsf\amsftrue
\ifx\bbbfont\nullfont
  \amsffalse
\fi
\ifamsf
\def\IR{\hbox{\bbbfont R}}
\def\IC{\hbox{\bbbfont C}}

\def\IZ{\hbox{\bbbfont Z}}

\def\antifund{\bar{\fund}}
\midinsert
\endinsert

\nref\giveon{A. Giveon and D. Kutasov, ``Brane Dynamics and Gauge Theory'',
hep-th/9802067.}

\nref\hw{A.~Hanany, E.~Witten, ``Type IIB Superstrings, BPS Monopoles and
Three-Dimensional Gauge Dynamics'', Nucl. Phys. B492 (1997) 152,
hep-th/9711230.}  

\nref\sixd{I.~Brunner, A.~Karch ``Branes and Six Dimensional Fixed
Points'', Phys. Lett. B409 (1997) 109-116, hep-th/9705022; ``Branes and
Orientifolds Versus Hanany-Witten in Six Dimensions'', JHEP 03(1998)003,
hep-th/9712143.
A.~Hanany, A.~Zaffaroni, ``Branes and Six-Dimensional Supersymmetric
Theories'', hep-th/9712145.}

\nref\seibergsix{N.~Seiberg, ``Non-trivial Fixed Points of
the Renormalization Group in Six Dimensions'', Phys. Lett.
B390 (1997) 169, hep-th/9609161.} 
\nref\othersix{K.~Intriligator, ``RG Fixed Points in Six Dimensions
Via Branes at Orbifold Singularities'', Nucl. Phys. B496 (1997) 177,
hep-th/9702038. J.~Blum, K.~Intriligator, ``Consistency Conditions for
Branes at Orbifold Singularities'', Nucl.Phys. B506 (1997) 223,
hep-th/9705030; ``New Phases of String Theory and 6d RG Fixed Points Via
Branes at Orbifold Singularities'', Nucl. Phys. B506 (1997) 199,
hep-th/9705044}

\nref\zaffa{ A. Hanany and A. Zaffaroni, ``On the Realization of Chiral
Four-dimensional Gauge Theories Using Branes'', JHEP 
05(1998)001, hep-th/9801134.}

\nref\barbon{J.~L.~F.~Barb\'on, ``Rotated Branes and $N=1$ Duality'',
Phys. Lett. B402 (1997) 59, hep-th/9703051.}

\nref\chiralfourd{K.~Landsteiner, E.~L\'opez, D.~A.~Lowe, ``Duality of
Chiral $N=1$ Supersymmetric Gauge Theories Via Branes'', JHEP 02(1998)007,
hep-th/9801002.
I.~Brunner, A.~Hanany, A.~Karch, D.~L\"ust, ``Brane Dynamics and Chiral
non-Chiral Transitions'', hep-th/9801017. S.~Elitzur, A.~Giveon,
D.~Kutasov, D.~Tsabar, ``Branes, Orientifolds and Chiral Gauge
Theories'', hep-th/9801020.}

\nref\lpt{J.~Lykken, E.~Poppitz, S.~P.~Trivedi, ``Chiral gauge theories
from D-branes'', Phys.Lett. B416(1998)286, hep-th/9708134; ``M(ore) on
chiral gauge theories from D-branes'', hep-th/9712193.}

\nref\math{A. Hanany, M. Strassler and A.M. Uranga, ``Finite Theories and
Marginal Operators on the Brane'', hep-th/9803086.}

\nref\bending{L.~Randall, Y.~Shirman, R.~von Unge, ``Brane Boxes: Bending
and the Beta Function'', hep-th/9806092.}

\nref\hanur{A. Hanany and A.M. Uranga, ``Brane Boxes and Branes on
Singularities'', hep-th/9805139.}

\nref\dm{M.R. Douglas and G. Moore, ``D-branes, Quivers and ALE
Instantons, hep-th/9603167.}

\nref\johnson{C.~V.~Johnson, R.~C.~Myers, ``Aspects of Type IIB Theory on
ALE Spaces'', Phys. Rev. D55 (1997) 6382, hep-th/9610140.}

\nref\mohri{ K. Mohri, ``D-branes and Quotients Singularities of Calabi-Yau
Fourfolds'', Nucl. Phys B521(1998)161, hep-th/9707012.}

\nref\dgm{ M.R. Douglas, B.R. Greene and D.R. Morrison, ``Orbifolds
Resolution by D-branes'', Nucl. Phys. B506(1997)84, hep-th/9704151.}

\nref\vafa{ A. Lawrence, N. Nekrasov and C. Vafa, ``On Conformal Field Theories
in Four Dimensions'', hep-th/9803015.}

\nref\zetatres{L.~E.~Ib\'a\~nez, ``A Chiral $D=4$, $N=1$ String Vacuum
with a Finite Low Energy Effective Field Theory'', hep-th/9802103.}

\nref\ks{S. Kachru and E. Silverstein, ``4d Conformal Field Theories and 
Strings on Orbifolds'', Phys. Rev. Lett. 80(1998)4855,
hep-th/9802183.}

\nref\angles{M.~Berkooz, M.~R.~Douglas, R.~G.~Leigh, ``Branes
Intersecting at Angles'', Nucl.Phys. B480 (1996) 265.}

\nref\hanhori{A. Hanany and K. Hori, ``Branes and $N=2$ Theories in Two
Dimensions'', Nucl. Phys. B513(1998)119, hep-th/9707192.}

\nref\fourfour{J.~H.~Brodie, ``Two Dimensional Mirror Symmetry From
M-theory'', Nucl.Phys. B517 (1998) 36, hep-th/9709228. M.~Alishahiha,
``N=(4,4) 2D Supersymmetric Gauge Theory and Brane Configuration'',
Phys. Lett. B420 (1998) 51, hep-th/9710020; ``On the Brane Configuration of
$N=(4,4)$ 2D Supersymmetric Gauge Theory'', hep-th/9802151.}

\nref\lerche{W.~Lerche, ``Fayet-Iliopoulos Potentials from Four-Folds'', 
JHEP 11(1998)004, hep-th/9709146.}

\nref\witten{E. Witten, ``Phases of $N=2$ Theories in Two Dimensions'',
Nucl. Phys. B403(1993)159, hep-th/9301042.}

\nref\distler{ J. Distler, ``Notes on (0,2) Superconformal Field Theories'',
hep-th/9502012.}

\nref\ed{E. Witten. ``Solutions of Four-dimensional Field Theories Via M
Theory'', Nucl. Phys. B500 (1997) 3.}

\nref\inst{E.~Witten, ``Sigma Models and the ADHM Construction of
Instantons'', J. Geom. Phys. 15 (1995) 215, hep-th/9410052.}

\nref\dasmuk{K.~Dasgupta, S.~Mukhi, ``A Note on Low-Dimensional String
Compactifications'', Phys. Lett. B398 (1997) 285, hep-th/9612188.}

\nref\bss{T. Banks, N.~Seiberg, E.~Silverstein, ``Zero and
One-dimensional Probes with $N=8$ Supersymmetry'', Phys. Lett.
 B401 (1997) 30, hep-th/9703052.}

\nref\agm{P.~S.~Aspinwall, B.~R.~Greene, D.~R.~Morrison, ``Calabi-Yau
Moduli Space, Mirror Manifolds and Spacetime Topology Change in String
Theory'', Nucl.Phys. B416(1994)414.}

\nref\flops{T.~Muto, ``D-branes on Orbifolds and Topology Change'',
hep-th/9711090; B.~R.~Greene, ``D-Brane Topology Changing Transitions'',
hep-th/9711124; S.~Mukhopadhyay, K.~Ray, ``Conifolds from D-branes'',
Phys. Lett. B423(1998)62.}

\nref\alwis{S.P. de Alwis, ``Coupling of Branes and Normalization of
Effective Actions in String/M-theory'',  Phys. Rev. 
D56(1997)7963, hep-th/9705139.}

\nref\douglas{ M.R. Douglas, ``Branes within Branes'', hep-th/9512077.}

\Title{hep-th/9806177, IASSNS-HEP-98/58}
{\vbox{\centerline{Brane Box Realization of Chiral}
\medskip
\centerline{Gauge Theories in Two Dimensions }}}
\smallskip
\centerline{ Hugo
 Garc\'{\i}a-Compe\'an\foot{Also {\it Departamento de
F\'{\i}sica, Centro de Investigaci\'on y de Estudios Avanzados del IPN,
Apdo. Postal 14-740, 07000, M\'exico D.F., M\'exico} E-mail:
compean@sns.ias.edu}, Angel M. Uranga\foot{E-mail: uranga@sns.ias.edu},}
\smallskip
\centerline{\it School of Natural Sciences}
\centerline{\it Institute for Advanced Study}
\centerline{\it Olden Lane, Princeton, NJ 08540, USA}
\vskip .5truecm

\bigskip
\medskip
\vskip .5truecm
\noindent

\centerline{\bf Abstract}
We study type IIA configurations of D4 branes and three kinds of NS
fivebranes. The D4 brane world-volume has finite extent in three
directions, giving rise to a two-dimensional low-energy field theory. The
models have generically $(0,2)$ supersymmetry. We determine the rules to
read off the spectrum and interactions of the field theory from the brane
box configuration data. We discuss the construction of theories with
enhanced $(0,4)$, $(0,6)$ and $(0,8)$ supersymmetry. Using T-duality along
the directions in which the D4 branes are finite, the configuration can be
mapped to D1 branes at $\IC^4/\Gamma$ singularities, with $\Gamma$ an
abelian subgroup of $SU(4)$. This provides a rederivation of the rules in
the brane box model. The enhancement of supersymmetry has a nice
geometrical interpretation in the singularity picture in terms of the
holonomy group of the four-fold singularity.

\vskip 1truecm
\noindent

\noindent

\Date{June, 1998}

\newsec{Introduction}

The dynamics of D-branes in certain configurations of intersecting branes
encodes many field-theoretical facts about supersymmetric theories in
several dimensions (for a review, see \giveon). Gauge theories in $p+1$
dimensions with
sixteen supercharges can be obtained as the world-volume theories of flat
infinite Dp-branes. In the context of theories with eight supersymmetries
in $p$ dimensions, it was shown in \hw\ that such theories can be realized
by considering  Dp-branes with a world-volume which is finite in one
direction, in which the D brane ends on NS fivebranes. For definiteness,
let us take such world-volume spanning $0,1,2,\ldots, (p-1)$, and
with finite extent along 6. The brane is suspended between NS
fivebranes spanning 012345. The low
energy theory in the non-compact dimensions of the D-brane is
$p$-dimensional. It is still a gauge theory, but the presence of the NS
branes breaks half of the supersymmetries, so eight supercharges remain.
This construction has been generalized in several directions, and has
yielded
the realization of a large family of models in several dimensions. This
setup has also been exploited to compute different exact quantum results
in these theories. For a review of such achievements, see \giveon.

A nice property of the interplay of field theories and configurations of
branes is that the intersections of branes
can sometimes support chiral zero modes. This opens the possibility of
studying chiral gauge theories using branes. The simplest such example is
provided by the realization of six-dimensional theories with eight
supersymmetries, which are chiral. These can be realized in the
setup described above by taking $p=6$, {\it i.e.} one considers D6 branes
extending along 0123456, and which are bounded in 6 by NS branes with
world-volume along 012345. This construction has
been discussed at length in \sixd, where it was shown that the
configurations yield theories satisfying the very restricting anomaly
cancellation conditions. The family of models obtained reproduces
the results of \refs{\seibergsix,\othersix}, and even contains some
further consistent examples.

Chirality if a fragile property, in the sense that toroidal
compactifications or too much supersymmetry spoil it. Thus, in order to
obtain chiral theories in four dimensions one has to consider theories with
only four supercharges. Their realization in terms
of branes requires new ingredients. A fairly general family of brane
configurations realizing generically chiral gauge theories in four
dimensions was introduced  in \zaffa \foot{Let us also mention that
the approach of rotated branes \barbon\ has been used to construct some
chiral gauge theories in four dimensions \chiralfourd. This
construction, however, seems not so flexible and easy to generalize. An 
interesting apporach, introduced in \lpt, is related to the 
constructions of \zaffa\ by T-duality.}. The 
idea is a clever extension of the
philosophy in \hw. It consists in realizing first a five-dimensional
theory with eight supercharges, by using D5 branes along 012346,
suspended between NS branes with world-volume along 012345. Then, the D5
brane is bounded
in the direction 4, by using a new set of NS branes oriented along 012367
(denoted NS$'$ branes). The low energy theory is four-dimensional, since
the world-volume of the D5 brane  along 46 is a finite rectangle. Such
configurations are known as brane box models. The presence
of the new kind of branes breaks a further half of the supersymmetries,
and so the theory has only four supercharges. Furthermore, the
intersections of NS, NS$'$ and D5 branes introduce chirality in the
four dimensional theory. There is no complete understanding of the
quantum effects of these gauge theories in terms of branes, even though
some results on exact finiteness and marginality were obtained in \math.
Some understanding on the bending of the branes in this configurations,
recently developed in \bending, may help in improving the situation in
this respect.

In this paper we continue this analysis of chirality in field theories
realized by brane configurations, and construct chiral gauge theories in
two dimensions, with two supercharges, {\it i.e.} $(0,2)$ theories. After
the previous discussion, there is a natural approach to the construction
of such theories. We first realize three-dimensional gauge theories with four
supercharges, by using D4 branes along 01246 and NS, NS$'$ branes as
before. The configuration is T-dual (along 3) to the brane box models
constructed above. We then bound the D4 branes in the direction 2,
by means of a new set of NS fivebranes, along 014567, denoted NS$''$ branes.
This implies the the low energy field theory will be two-dimensional, and
that only two supersymmetries remain unbroken. Now it
is the intersection of D4, NS, NS$'$ and NS$''$ branes that introduces the
chirality in the two dimensional theory. The construction and study (at
the classical level) of these theories is the aim of the present
paper.

We start in Section~2 by reviewing some basic features of two-dimensional
field theories. In Section~3 we describe the brane configurations sketched
above. The spectrum and interactions corresponding to a given brane model
is determined by
studying first the realization of $(2,2)$ theories, and generalizing the
result.

In Section~4 we discuss models with enhanced supersymmetry.
An interesting feature of two-dimensional theories is that it is possible
to enhance the supersymmetry while preserving chirality. So it makes sense
to ask whether new ingredients are required to realize {\it e.g.} $(0,4)$
theories using brane constructions. In Section~4 we show a simple way of
realizing a large class of models with such enhanced chiral supersymmetry
in our setup, without the need of extra ingredients.

In Section~5 we discuss the relation or our approach to that of \mohri.
Recently it has become clear that there is an
alternative way of realizing chiral theories in several dimensions, as
the world-volume theories of D brane probes at singularities. Thus, using
D5 branes at ADE singularities (possibly with some orientifold projection)
one can realize six-dimensional theories
with eight supercharges \refs{\dm,\johnson,\othersix}. Some of these
constructions (corresponding to $A_k$ singularities) have
been argued to be T-dual to the brane configurations in \sixd.
Similarly, by using D3 branes at $\IC^3/\Gamma$ singularities, with $\Gamma$ a
discrete subgroup of $SU(3)$, one can realize chiral four-dimensional theories
with four supercharges \refs{\dgm,\vafa,\zetatres,\ks}. For abelian
discrete
groups, these constructions are related by T-duality to the brane box
models, as argued in \hanur. In Section~5 we apply the same argument to
the our brane models, and
discuss their relation with the theories on the world-volume of D1 branes
at $\IC^4/\Gamma$ singularities, with $\Gamma$ a discrete subgroup of
$SU(4)$. These latter theories have been studied in \mohri, and we review
the determination of the spectrum and interactions in the singularity
language. This can be compared with the rules proposed in Section~3, and
provide a rederivation of the result. Also, the agreement supports our
T-duality proposal in this case. Finally, it
provides a nice geometrical interpretation for the enhancement to $(0,4)$
and $(0,8)$ supersymmetry introduced in Section~4. The explanation
suggests the brane configurations with enhanced supersymmetry have an
appropriate `generalized holonomy group' in the sense of \angles.

Finally, Section~6 contains our conclusions. The definition of the gauge
theory parameters in terms of branes is discussed in an appendix.

We note that brane configurations have been used to obtain
several results on non-chiral theories in two dimensions, in \hanhori\ for
$(2,2)$ theories, and in \fourfour\ for $(4,4)$ theories. Another approach
to the study of $(2,2)$ theories has been taken in \lerche\ in the
framework of geometric engineering.

\vskip 2truecm


\newsec{Overview of Field Theory}

In this section we will give an overview of the matter content and
interactions of the two-dimensional (2,2) and (0,2) gauge theories. Our
aim is not to provide an extensive review of such
theories, but to briefly recall the relevant structure of multiplets
and interactions, which we will need in the following sections. For a more
complete treatment see Refs. \refs{\witten,\distler}.

\vskip 1truecm

\subsec{$\NN=1$ Theories in Four Dimensions}

Let us first briefly review the matter content and interactions of
${\cal N}=1$ gauge theories in four dimensions. For definiteness
let us take the gauge group to be $U(N)$. The theory is
formulated in the superspace consisting of four spacetime coordinates
$x^{\mu}$, ($\mu = 0,1,2,3$) and odd coordinates $\theta^{\alpha}$,
$\bar{\theta}^{\dot{\alpha}}$
The multiplets used in the construction of gauge theories are:

{\it i)-.} The {\it chiral multiplet}, which we will denote by
$\tilde{\Phi}$,
which contains a complex scalar and a chiral fermion,
in any representation of the gauge group.

{\it ii)-.}The {\it gauge multiplet}, $\tilde{V}$, containing (in the
Wess-Zumino gauge) gauge bosons and a
Majorana fermion, both in the adjoint representation of the gauge group.

The interactions of the ${\cal N}=1$ theory are encoded in the superspace
Lagrangian

$$L = \int d^4x\, d^4\theta \sum_i \bigg(\bar{\tilde{\Phi}}_i
 \exp\big({{\sum}_a T_a \tilde{V}_a}\big) \tilde{\Phi}_i\bigg) + \int
d^4x\, d^2 \theta\, {\rm Tr} (W^{\alpha}W_{\alpha})$$
\eqn\lagrfourd{ + \int d^4x\, d^2 \theta \ W(\tilde{\Phi}_i) + h.c.
 \hfill {} - r \int d^4x\, d^4 \theta \ {\rm Tr}(\tilde{V})}
where $W_{\alpha} = -{1\over 4} \bar{D}^2 D_{\alpha} \tilde{V}$ is the
gauge field strength chiral superfield. The first term
contains the kinetic term and gauge couplings for the chiral multiplets.
Here $T_a$ are the generators of the gauge group in the appropriate
representation. The second term is the kinetic energy and
gauge interactions for the gauge
multiplets. The third term is the superpotential; $W(\tilde{\Phi}_i)$ is
a
holomorphic function of the chiral superfields ${\tilde \Phi_i}$. In the
applications we are to consider, the superpotential will be cubic, and
contains the Yukawa interactions and scalar potential terms. The last term
is the Fayet-Iliopoulos term which arises if there is a U(1)
factor in the gauge group. Finally let us mention that
${\cal N}=1$ gauge theories have a U(1) R-symmetry. This symmetry can be
broken to a discrete subgroup by instanton effects.

\vskip 1truecm
\subsec{(2,2) Theories in Two Dimensions}

We now describe how dimensional reduction of ${\cal N}=1$ in four
dimensions yields (2,2) gauge theories in two dimensions. To this end, one
takes all fields to be independent of $x^2,x^3$, and decomposes the
representations of the four-dimensional Lorentz group with respect to the
two-dimensional one. The resulting gauge theory in two dimensions is
nicely described in the (2,2) superspace
$(y^{\alpha},\theta^+,\theta^-,\bar{\theta}^+,\bar{\theta}^-)$, where
$(y^0,y^1)=(x^0,x^1)$.

The structure of multiplets is very similar to the four-dimensional one,
as follows.

{\it i)-.} There is a (2,2) {\it chiral multiplet} ${\Phi}$,
containing a complex
scalar $\phi$, and two fermions of opposite chirality $\psi_+$, $\psi_-$.

{\it ii)-.} There is also a {\it vector multiplet} ${V}$, containing
gauge bosons, $v_{\alpha}$, $\alpha=0,1$,
two Majorana fermions $\lambda_+$, $\lambda_-$ and one complex scalar
$\sigma$ (this last arising from the components of the four-dimensional
vector along $x^2,x^3$).

The Lagrangian is obtained by simple reduction of the Lagrangian
\lagrfourd. It has the following structure,
\eqn\dosdim{L= L_{ch} + L_{gauge} + L_W + L_{D,\theta}.}

The term $L_{ch}$ is the kinetic energy and gauge couplings of the chiral
superfields $\Phi$

\eqn\interaccion{ \int d^2\, y d^4 \theta \sum_i \bigg(\bar{\Phi}_{i}
exp\big({\sum}_a T^a V_a \big) \Phi_{i}\bigg).}

The term $L_{gauge}$ contains the kinetic energy and gauge interactions
for the vector multiplets, and takes the form

\eqn\gautwotwo{ L_{gauge} = - {1\over{g^2}}\int d^2y \, d^4\theta
 {\rm Tr} \big({\bar \Sigma}\Sigma\big),}
where $\Sigma$ is the $(2,2)$ gauge field strength superfield and $g$ is
the gauge coupling constant.

The $(2,2)$ superpotential term $L_W$ is given by

\eqn\ldos{L_W = - \int d^2 y\, d\theta^+ d \theta^- W(\Phi_i)
|_{\bar{\theta}^+=\bar{\theta}^-=0} - h.c. \ .}

Finally, the last term contains the Fayet-Iliopoulos  and theta angle
terms,

\eqn\ltres{L_{D,\theta} = {it\over 2 \sqrt{2}} \int  d^2y d\theta^+
d \bar{\theta}^-  {\rm Tr}\big(\Sigma
|_{\theta^- = \bar{\theta}^+ =0} \big) + h.c. }
where $t = ir + {\theta \over 2 \pi}$.

Among the symmetries present in (2,2) theories, the R-symmetries play an
important role. For these theories there can two U(1) R-symmetries, the
right-moving
U(1) R-symmetry acts on the right-moving odd coordinates $(\theta^+,
\bar{\theta}^+)$ in the form $\theta^+ \to e^{i \beta} \theta^+$,
$\bar{\theta}^+ \to e^{-i \beta} \bar{\theta}^+$, leaving $\theta_-$  and
$\bar{\theta}^-$ invariant. Similar definition holds for the left-moving
U(1) R-symmetry acting on the left-moving $(\theta^-,
\bar{\theta}^-)$. Since the fermions are charged under these symmetries
they can be anomalous. The condition that the mixed anomaly of the
R-symmetries and the $U(1)$ gauge factor in $U(N)$ vanishes is

\eqn\anomaly{ \sum_i Q_{i} = 0,}
where $Q_{i}$ is the gauge $U(1)$ charge of the $i$-th chiral field
$\Phi_i$. This ensures the conservation of the left-moving and
right-moving R-currents $J_L$, $J_R$. This relation also implies the
non-renormalization of the coefficient of the Fayet-Iliopoulos term.

\vskip 1truecm
\subsec{(0,2) Theories in Two Dimensions}

We now move on to review the building blocks of (0,2) gauge theories
in two dimensions. These theories are described in the (0,2)
superspace $(y^{\alpha},\theta^+,\bar{\theta}^+)$. There are three basic
kinds of multiplets which we will use.

{\it i)-.} The $(0,2)$ {\it gauge multiplet} $V'$, which contains gauge bosons
$v_{\alpha}$, $\alpha=0,1$, and one fermion $\chi_-$.

{\it ii)-.} The $(0,2)$ {\it chiral multiplet} $\Phi'$, contains one complex
scalar $\phi$ and one chiral fermion $\psi_+$.

{\it iii)-.} The $(0,2)$ {\it Fermi multiplet}, $\Lambda$, is described
by an anticommuting superfield. Its complete $\theta$ expansion contains a
chiral spinor $\lambda_-$,
an auxiliary field $G$, and a holomorphic function $E$ depending on the
chiral (0,2) superfields $\Phi_i'$. The Fermi multiplet  $\Lambda$
satisfies the constraint $\bar{\cal D}_+ \Lambda= \sqrt{2} E(\Phi')$,
with $\bar{\cal D}_+ E = 0$. Here $\bar{\cal D}_+$ represents the supersymmetric
covariant derivative. The expansion in components for the Fermi superfield
is
\eqn\fermi{\Lambda=\lambda_- -\sqrt{2}\theta^+ G - i\theta^+ {\bar
\theta}^+ (D_0+D_1)\lambda_- - \sqrt{2} {\bar \theta}^+ E(\Phi')}
with $D_{\alpha}$ denoting the usual supersymmetric derivative 
\foot{Notice that our definitions of the Fermi multiplets and the 
function E follow \witten, and differ from those in \distler.}.

Gauge theories involving these fields are described by a Lagrangian with
the following structure,
\eqn\lcuatro{L =  L_{gauge} + L_{ch} + L_F + L_{D,\theta} + L_J .}
As usual, $L_{gauge}$ is the kinetic term of the gauge multiplet given by
\eqn\uno{ L_{gauge} = {1 \over 8g^2} \int d^2y d\theta^+ d \bar{\theta}^+
{\rm Tr} \big(\bar{\Upsilon} \Upsilon\big)}
where $\Upsilon$ is the field strength of $V'$.

The term $L_{ch}$ contains the
kinetic energy and gauge couplings of the (0,2) chiral superfields
$\Phi'_i$. It is given by

\eqn\dos{L_{ch} = - {i \over 2} \int d^2y d^2 \theta \sum_i
\bigg(\bar{\Phi'_i}({\cal D}_0 - {\cal D}_1) \Phi'_i \bigg),}
where ${\cal D}_{0}$ and ${\cal D}_{1}$ are the (0,2) gauge covariant derivatives
with respect to $V'$.

The term $L_F$ describes the dynamics of the Fermi multiplets $\Lambda$,
and certain interactions. It is given by

\eqn\tres{L_F= -{1\over 2} \int d^2yd^2\theta \sum_a \big(
\bar{\Lambda}_{a} {\Lambda}_{a} \big).}

For future convenience, we give the expression of this term in components,
as obtained upon substitution of \fermi\ in \tres.

\eqn\lagrferm{L_F\, =\, \int d^2y \sum_a \bigg\{
i{\bar\lambda_{-,a}}(D_0+D_1)
\lambda_{-,a} + |G_a|^2 - |E_a|^2 - \sum_j\big( {\bar \lambda_{-,a}}
{\partial E_a
\over \partial \phi_j}\, \psi_{+,j} + {\partial \bar{E_a} \over \partial
\bar{\phi_j}}\, {\bar \psi_{+,j}} \lambda_{-,a} \big) \bigg\}.}

The Fayet-Iliopoulos and theta angle terms are encoded in the (0,2)
Lagrangian $L_{D,\theta}$ which is written as
\eqn\cuatro{ L_{D,\theta} = {t \over 4} \int d^2 y d\theta^+ {\rm Tr}
\big(\Upsilon
|_{\bar{\theta}^+ = 0} \big) + h.c. }
where $t = {\theta\over 2 \pi} + ir$.

Finally  (0,2)  models do admit an additional
interaction term $L_J$ which depends on a set of holomorphic functions
$J^a(\Phi')$ of the chiral superfields. There is one such function for
each Fermi
superfield. They satisfy the relation $\sum_a E_a J^a =0$.
This interaction is the (0,2) analog of the superpotential, and its
Lagrangian $L_J$ is given by

 \eqn\cinco{ L_J = -{1 \over \sqrt{2}} \int d^2 y d\theta^+
 \sum_a \bigg(\Lambda_{a}J^a|_{\bar{\theta}^+ = 0} \bigg) - h.c. \ \ .}

The expansion of this term in components is

\eqn\lagrj{L_J=-\int d^2y \sum_a \bigg( G_a J^a + \sum_j \lambda_{-,a}
\psi_{+,j}
{{\partial J^a} \over {\partial \phi_j}} \bigg) - h.c. \ \ .}

After combining the Lagrangians $L_F$ and $L_J$ and
solving for the equations of motion for the auxiliary fields $G$,
the relevant interaction terms in the Lagrangian (we are not listing
the gauge interactions and D-terms here) are

\eqn\interlagr{\sum_a \big( |J^a(\phi)|^2 + |E^a(\phi)|^2 \big)
- \sum_{a,j} \big( {\bar \lambda_{-,a}} {\partial E_a \over \partial
\phi_j} \psi_{+,j} + \lambda_a {\partial J^a \over \partial \phi_j}
\psi_{+,j} + h.c.\big). }

The first term contains the scalar potential, and the second the Yukawa
couplings. Notice that the choice of the functions $E$ and $J$ completely
defines the interactions of the theory.

For (0,2) theories in two dimensions we have just one U(1) R-symmetry
group acting on the superspace coordinates $(\theta^+,\bar{\theta}^+)$.
This is right-moving  R-symmetry and it acts as
$\theta^+ \to e^{i \beta} \theta^+$,
$\bar{\theta}^+ \to e^{-i \beta} \bar{\theta}^+$, leaving $\theta^-$ ,
$\bar{\theta}^-$ invariant. Again, Eq. \anomaly\ provides the condition to
have cancellation of the mixed anomalies.

\vskip 1truecm
\subsec{Decomposition of (2,2) Superfields In Terms of (0,2) Multiplets}

For future convenience it will be useful to decompose the $(2,2)$ theories
in terms of (0,2) superfields. Looking at the field content of the
multiplets, we conclude that
the $(2,2)$ vector multiplet $V$ decomposes as a (0,2) gauge
multiplet $V'$ (containing the gauge boson $v_{\alpha}$ ($\alpha =0,1$)
and the negative chirality spinor $\chi_-$) and a (0,2) chiral
multiplet  $\Sigma'= \Sigma|_{\theta^-=\bar{\theta}^-=0}$ consisting of an
scalar $\sigma$ and the positive chirality spinor $\chi_+$
in the adjoint representation of the gauge group.

As for the (2,2) chiral multiplet $\Phi_i$, it decomposes as a (0,2)
chiral
multiplet
$\Phi'_i= \Phi_i|_{\theta^-= \bar{\theta}^- =0}$  (containing the complex
scalar and $\phi_i$ and the positive chirality spinor $\psi_{+,i}$) and a
(0,2)
Fermi multiplet  $\Lambda_{i}$ (containing the negative chirality spinor
$\psi_{-,i}$, denoted $\lambda_{-,i}$ in what follows in order to be
consistent with our $(0,2)$ conventions).

This Fermi multiplet is given by $\Lambda_{i} =
{1 \over \sqrt{2}} {\cal D}_-
\Phi_i|_{\theta^-= \bar{\theta}^- =0}$. It can be verified that the
corresponding function $E_i$ of the $(0,2)$ chiral superfields is
$E_i = i\sqrt{2} T_a \Sigma'_a \Phi_i'$, where $a$ runs
over the generators of the gauge group under which the field is charged,
and $T_a$ are in the appropriate representation.

Some of the interactions in the $(0,2)$ theory are basically specified by
the gauge group and the representations of the chiral and Fermi
multiplets. Only the $L_J$ term and the interactions coming from $L_F$
deserve special discussion. The Lagrangian $L_J$ is obtained by
reduction of the $(2,2)$ superpotential, and thus involves the chiral and
Fermi fields coming from the $(2,2)$ chiral fields. The Lagrangian takes
the the form \cinco\ with a specific form for the functions $J^i$, namely
\eqn\seis{ J^i = {\partial W \over \partial \Phi'_i}}
with $W$ being the superpotential of the (2,2) theory.
The equation $\sum_i E_i J^i=0$ follows from gauge invariance of $W$.

There are also interactions between the $(0,2)$ chiral multiplet $\Sigma'$
and the chiral and Fermi multiplets $\Phi'_i$, $\Lambda_i$.
These couplings are gauge interactions from the $(2,2)$
point of view, but in the $(0,2)$ theory appear from the expansion of
$L_F$. They are obtained by substituting in Eq.\lagrferm\ the
functions $E_i$ found above.

\vskip 2truecm


\newsec{The Brane Configurations}

In this section we introduce certain supersymmetric configurations of
NS, NS$'$, and NS$''$ branes, and D4 branes in Type IIA superstring
theory. They give rise to two-dimensional $(0,2)$
field theories. These configurations are obtained
in the spirit of the brane box configurations in \zaffa, by considering
D-branes which are finite in several directions. As explained in the
introduction, they belong to a natural sequence of brane box models
yielding chiral theories in six, four and two dimensions (taking D branes
compact in one, two and three directions, respectively).

\vskip 1truecm
\subsec{Description of the Brane Configurations}

Let us consider the ingredients of the brane configurations which we will
use in this paper. Brane configurations consist of:

\item{a)-.} NS fivebranes located  along $(012345)$.

\item{b)-.} NS' fivebranes located  along $(012367)$.

\item{c)-.} NS'' fivebranes located  along $(014567)$.

\item{d)-.} D4 branes located along $(01246)$.

In this configuration the D4 branes are finite in the directions
246. They are bounded in the direction $2$ by the NS$''$ branes, in
the direction $4$ by the NS$'$ branes, and in the direction $6$ by the NS
branes. For the D4 branes to be suspended in this way, it is necessary
that the coordinates of all branes in $89$ should be equal. It is also
required that  two NS branes joined by a D4 brane should have the
same position in 7, and analogously that two NS$'$ branes joined by a
D4 brane should have the same
position in 5, and that two NS$''$ branes should have the same
position in 3. In Figure~1 we show the $246$ three-dimensional space
and illustrate how the NS, NS$'$, and NS$''$ brane bound the D4 brane in
three directions.

The low-energy effective field theory on the D4 branes is two-dimensional,
since $01$ are the only non-compact directions in their world-volume. The
presence of each kind
of NS fivebrane breaks one half of the supersymmetries, and altogether
they break to $1/8$ of the original supersymmetry. A further half is
broken by the D4 branes, and the world-volume theory has
$(0,2)$ supersymmetry in two dimensions. Since the D brane is bounded by
NS fivebranes, the world-volume gauge bosons will not be projected out and
there will be a gauge group for each box in the model.
The $U(1)_R$ R-symmetry of the field theory is manifest as the rotational
symmetry in the directions 89.

We note that there are a variety of other objects that can be introduced
in the configuration without breaking the supersymmetry. For instance,
there are three kinds of D6 branes that can be introduced, namely D6
branes along 0124789, D6$'$ branes along 0125689, and D6$''$ branes along
0134689. They provide vector-like flavours for the gauge groups. These
extensions are quite well-known from other contexts, and we will not study
them in the present paper.

\bigskip
\centerline{\vbox{\hsize=4in\tenpoint
\centerline{\psfig{figure=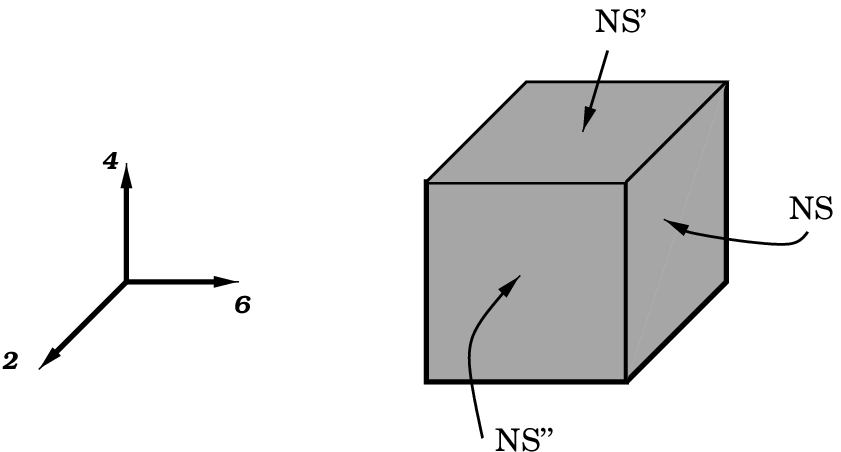}}
\vglue .2in
{\ninerm Fig. 1.  A three-dimensional box with 2 NS branes, 2 NS$'$
branes and
2 NS$''$ branes. The D4 brane world-volume fills the cube bounded by the
NS fivebranes. For clarity we have shown only a finite interval of the
NS, NS$'$ and NS$''$ branes, which actually extend to infinity along
24, 26 and 46, respectively.}}}

There is a first rough classification we can make in these brane
configurations, according to whether the directions 246 are taken compact
or not. If some of these directions are non-compact, then there will be
some semi-infinite box, which will represent some global symmetry. For
definiteness we will center on the case in which all three directions are
compact, with lengths $R_2$, $R_4$ and $R_6$. Extension of our results to
other cases is straightforward.

A generic configuration consists of a three-dimensional grid of $k$ NS
branes, $k'$ NS' branes and $k''$ NS'' branes dividing the 246
torus into a set of $kk'k''$ boxes. We will often think about these
configurations as infinite periodic arrays of boxes in $\IR^3$,
quotiented by an infinite discrete group of translations in a
three-dimensional lattice $\Lambda$. This point
of view is particularly useful to define models in which the unit cell has
non-trivial identifications of sides \refs{\math,\hanur}. In Section~4 we
will present some examples of this last case.

\vskip 1truecm

\subsec{Motivating the Brane Box Rules}

The next question we would like to address is what is the spectrum of the
$(0,2)$ gauge theory corresponding to a given brane box configuration.
The strategy we are to follow in this section is to first consider a
particular family of brane box models, namely those with $k''=1$. Notice
that if there are no NS$''$ present, the brane configurations can be
thought of as a compactification of the brane box models in
\refs{\zaffa,\math}, up to a T-duality along direction 3. The
corresponding two-dimensional field theory will be a dimensional reduction
of the four-dimensional $\NN=1$ gauge theories
in \zaffa, thus a $(2,2)$ theory. The spectrum and interactions of
such theory are known from the four-dimensional analysis, where it was
shown that fields transforming in bi-fundamental representations could
be represented as arrows joining the D-branes in different boxes
(representing the open strings stretched between the D branes), and
superpotential interactions corresponded to closed triangles of arrows
(representing open string interactions).
Our purpose is to reinterpret such fields and interactions in the $(0,2)$
language, and find some rules yielding the correct spectrum. These will
turn out to have a natural generalization to other brane configurations
with several NS$''$ branes, and which yield genuine $(0,2)$ field
theories.

Let us mention that the introduction of a {\it single}
NS$''$ brane does not change the theory, so it still has $(2,2)$
supersymmetry. It will be useful to introduce such brane, because then the
matter multiplets will appear in a suggestive pattern, easy to generalize
to $(0,2)$ models.

So let us consider a $k \times k' \times 1$ box model,
with trivial identifications of the sides of the unit cell, such
as that depicted in Figure~2 for the particular case $k=3$, $k'=4$. Let us
also place $n_{a,b}$
D4 branes in the box in the position $(a,b)$ in the grid.

\bigskip
\centerline{\vbox{\hsize=4in\tenpoint
\centerline{\psfig{figure=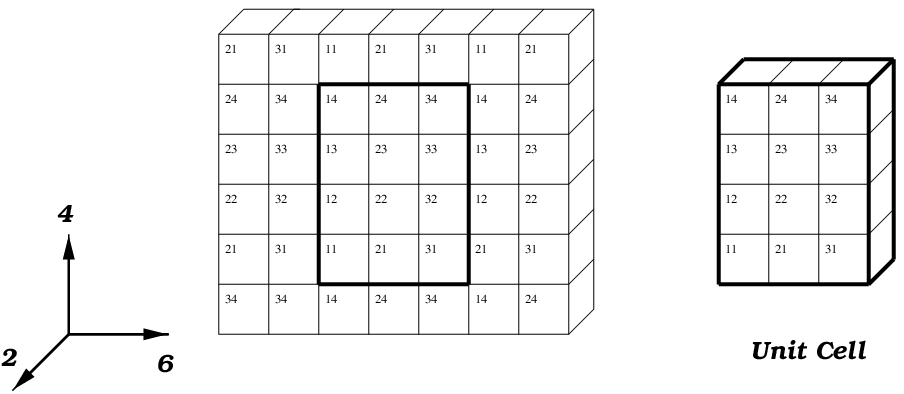}}
\vglue .2in
{\ninerm Fig. 2. A $3\times 4\times 1$ box model, with its twelve
different boxes and its unit cell. The sides of the unit cell are
identified to make the direction 246 compact. Equivalently, the brane
configuration can be described as an infinite periodic array of unit
cells. For clarity we only show one slice of the infinite periodic array
along~2. Notice that, in this and the following figures, the numbers in
the boxes are labels to distinguish
them and do not denote the number of D4 branes in the box.}}}

It will be convenient to imagine the brane configuration as an infinite
periodic array of boxes extending in 246. Thus, each box will be labeled
by three indices $(a,b,c)$ corresponding to its position in the grid in
6, 4 and 2, respectively. Even though things will not depend on the
actual value of $c$ (since the periodicity in 2 is of one box), again it
will be useful to maintain such notation, aiming to a
further generalization to be completed in following sections.

Using the $(2,2)$ reduction of the rules in \zaffa, we know that the
box $(a,b)$ has vector multiplets giving a $U(n_{a,b})$
gauge group\foot{In two dimensions we expect the $U(1)$ factors to
be dynamical. The freezing argument in \ed\ applies to theories in
four or more dimensions.}. There are also $(2,2)$ chiral multiplets which
corresponded
to horizontal, vertical and diagonal arrows in the four-dimensional
construction. We denote these fields by ${\Phi^H}_{a,b}$ (in the
$(\fund,\antifund)$ of $U(n_{a,b})\times U(n_{a+1,b})$), ${\Phi^V}_{a,b}$
(in the $(\fund,\antifund)$ of $U(n_{a,b})\times U(n_{a,b+1})$), and
$\Phi^D_{a,b}$ (in the $(\fund,\antifund)$ of $U(n_{a,b})\times
U(n_{a-1,b-1})$). There are three such fields for each box in the model.

Let us decompose this field content with respect to $(0,2)$ multiplets and
try to define some arrows in the brane box diagram to represent such
fields. As reviewed in section 2.4, the $(2,2)$ vector multiplet at
each box gives a $(0,2)$ vector multiplet and a $(0,2)$ chiral
multiplet in the adjoint of $U(n_{a,b})$. We can represent this field by
an arrow which starts in the box $(a,b,c)$, extends along 2, and ends in
another copy of the box $(a,b,c+1)$. This transforms in the adjoint, since
in this configuration the boxes that differ only in $c$ are identified due
to the periodicity in $2$. We will denote this $(0,2)$ field by
$N_{a,b,c}$.

The $(2,2)$ chiral field $\Phi^H_{a,b}$ gives rise to a $(0,2)$ chiral
field and a $(0,2)$ Fermi multiplet transforming in the
$(\fund,\antifund)$  of $U(n_{a,b})\times U(n_{a+1,b})$.
Our proposal is that the chiral
multiplet, denoted $H_{a,b,c}$, should be represented by an arrow which
goes from the box $(a,b,c)$ to the box $(a+1,b,c)$. The Fermi multiplet,
denoted $\Lambda^{(1)}_{a,b,c}$,
is however represented by an arrow starting from the box $(a,b,c)$ and
ending on the box $(a+1,b,c+1)$. It will be useful to also consider an
arrow starting from the box $(a+1,b,c+1)$ and ending on the box $(a,b,c)$,
which represents the conjugate of the Fermi multiplet, denoted
$\Lambda^{({\bar 1})}_{a+1,b,c+1}$. Notice however that this is not an
independent field \foot{Even though the notation may appear confusing,
we stick to denoting the fields by the box their arrow starts at.}. Notice
how a four-dimensional field splits in two due to the presence of the new
kind of NS fivebrane. This is somewhat analogous to how four-dimensional
$\NN=2$ theories are constructed using brane box models \math.

The $(2,2)$ chiral multiplet $\Phi^V_{a,b}$ analogously yields one $(0,2)$
chiral
multiplet and one Fermi multiplet transforming in the $(\fund,\antifund)$
of $U(n_{a,b})\times U(n_{a,b+1})$. We can represent the chiral multiplet,
denoted $V_{a,b,c}$ by an arrow going from the box $(a,b,c)$ to the box
$(a,b+1,c)$. The Fermi field denoted $\Lambda^{(2)}_{a,b,c}$  corresponds to
an arrow from the box $(a,b,c)$ to the box $(a,b+1,c+1)$. Again we
introduce another arrow, going from the box $(a,b+1,c+1)$ to the box
$(a,b,c)$, which represents the conjugate $\Lambda^{({\bar 2})}_{a,b+1,c+1}$
of the Fermi multiplet.

Finally, the $(2,2)$ chiral multiplet $\Phi^D_{a,b}$ gives rise to one
$(0,2)$ chiral multiplet and one Fermi multiplet transforming in the
$(\fund,\antifund)$ of $U(n_{a,b})\times U(n_{a-1,b-1})$. The chiral
multiplet, denoted $D_{a,b,c}$, is represented by an arrow going from the
box $(a,b,c)$ to the box $(a-1,b-1,c-1)$. The Fermi field, denoted
$\Lambda^{(3)}_{a,b,c}$, corresponds to an arrow
from the box $(a,b,c)$ to the box $(a-1,b-1,c)$. Again, the opposite arrow
is taken to represent the conjugate $\Lambda^{({\bar 3})}_{a-1,b-1,c}$.

The arrows defining the $(0,2)$ chiral fields are shown in Figure ~4,
and those for Fermi multiplets (and their conjugates) in Figure ~5. The list
of $(0,2)$ multiplets is given in Table~1, along with their origin in the
$(2,2)$ theory.

\bigskip
\centerline{\vbox{\hsize=4in\tenpoint
\centerline{\psfig{figure=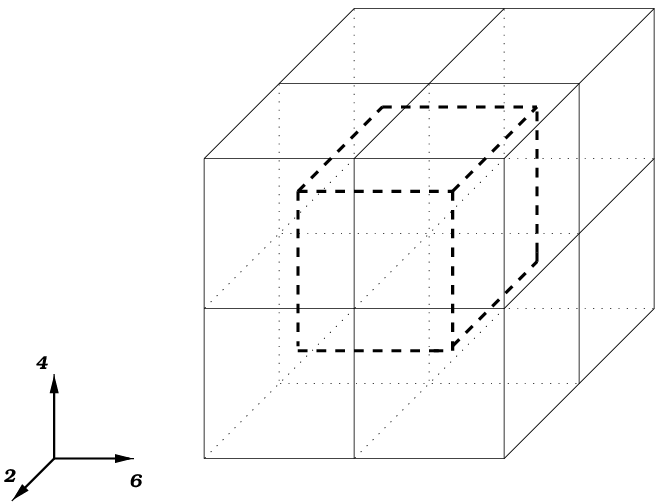}}
\vglue .2in
{\ninerm Fig. 3. The vertices of the cube with dashed contour
are located at the centers of the cube drawn with continuous lines.}}}

\bigskip
\centerline{\vbox{\hsize=4in\tenpoint
\centerline{\psfig{figure=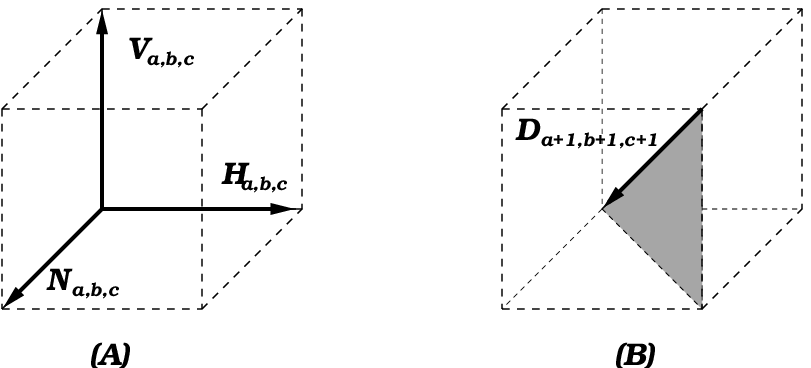}}
\vglue .2in
{\ninerm Fig. 4. Arrows representing the (0,2) chiral multiplets. For
clarity
we show the fields $H_{a,b,c}$, $V_{a,b,c}$ and $N_{a,b,c}$, and the
field $D_{a+1,b+1,c+1}$ in separate pictures. In this and following
pictures, the cube with dashed contour is as defined in Figure~3.}}}


\midinsert{
{\sevenrm
$$
\vbox{\offinterlineskip
\def\strut{\vrule height 3.25ex  width 0pt depth 2ex}
\def\hline{\noalign{\hrule}}
\halign{\vrule#\hfil\strut&\quad#\hfil\quad\vrule&&\quad\enspace
\hfil#\hfil\quad\vrule\cr
\noalign{\hrule}
&$\hfil{\bf (0,2)\;\; Field}$& ${\bf Representation}$ & ${\bf (2,2)\;\;
Field}$\cr
\noalign{\hrule}
\hline
&$\matrix{H_{a,b,c} \cr \Lambda^{(1)}_{a,b,c}}$& $\matrix{(\fund_{a,b,c},
\antifund_{a+1,b,c})\cr (\fund_{a,b,c},\antifund_{a+1,b,c+1})}$ &
$\matrix{\Phi^H_{a,b}}$ \cr
\hline
& $\matrix{V_{a,b,c}\cr \Lambda^{(2)}_{a,b,c}}$ & $\matrix{(\fund_{a,b,c},
\antifund_{a,b+1,c})\cr (\fund_{a,b,c},\antifund_{a,b+1,c+1})}$ &
$\matrix{\Phi^V_{a,b}}$\cr
\hline
& $\matrix{D_{a,b,c}\cr \Lambda^{(3)}_{a,b,c}}$& $\matrix{(\fund_{a,b,c},
\antifund_{a-1,b-1,c-1})\cr (\fund_{a,b,c},\antifund_{a-1,b-1,c})}$ &
$\matrix{\Phi^D_{a,b}}$\cr
\hline
& $N_{a,b,c}$& $(\fund_{a,b,c},\antifund_{a,b,c+1})$ (${\rm Adj.}$) &
${\rm Vector}$ \cr
\hline
\hline
}}
$$ }}

{\ninerm {Table 1: Fields in a generic (2,2) model obtained by reduction
of a four-dimensional $\NN=1$ model. The matter content is split in
(0,2) chiral and Fermi multiplets. The origin of the fields in the (2,2)
theory is specified in the last column.}}
\endinsert

Since the boxes are repeated periodically along 2 with period of one box, there
is some
ambiguity in the assignation of arrows. Now we show that with the above
choice the $(0,2)$ interactions arise from closed
triangles of arrows.

For each box there are six closed triangles involving the arrows
corresponding to the fields $H$, $V$, $D$ and $\Lambda^{(i)}$, $i=1,2,3$.
They are shown in figures 6C, 6D and 6E (The meaning of the other
triangles will be explained later). Recall that the $(0,2)$
superpotential interactions
involve one Fermi multiplet and two chiral multiplets, a feature
correctly
reproduced by our choice of arrows. As mentioned in section 2.4, the
$(0,2)$ interactions in $L_J$ arise from the reduction of the $(2,2)$
superpotential. Specifically, we expect an interaction of type \cinco\
involving the $(0,2)$ chiral and Fermi fields arising from $(2,2)$ chiral
multiplets, namely $H_{a,b,c}$, $V_{a,b,c}$ and $D_{a,b,c}$, and
$\Lambda^{(i)}_{a,b,c}$ $(i =1,2,3)$. The functions $J$ in that equation
can be obtained from the superpotential through Eq. \seis.
It is easy to check that these are given by the mentioned triangles.
Explicitly, we obtain the following $J$ functions:

\eqn\jfunctone{\matrix{
J^{(1)}_{a,b,c}\, & = & \, V_{a+1,b,c+1}D_{a+1,b+1,c+1}\, & - &\,
D_{a+1,b,c+1} V_{a,b-1,c} \cr
J^{(2)}_{a,b,c}\, &= & \, D_{a,b+1,c+1} H_{a-1,b,c}\, & - &
\, H_{a,b+1,c+1} D_{a+1,b+1,c+1}\cr
J^{(3)}_{a+1,b+1,c}\, & = & \,H_{a,b,c}V_{a+1,b,c}\, & -&\,
V_{a,b,c}H_{a,b+1,c}. \cr}}

The triangles can be thought of as describing the Yukawa interactions that
arise from the $(0,2)$ superpotential, as in \lagrj.

The remaining triangles contain the information about the $(0,2)$
interaction terms coming from the Lagrangian $L_F$, as in \lagrferm. These
are defined by the functions $E$ associated to the Fermi multiplets. These
can be obtained from the $(2,2)$ gauge interactions. The resulting
functions are

\eqn\efunctone{\matrix{
E^{(1)}_{a,b,c} \, & = &\, N_{a,b,c} H_{a,b,c+1}\, & -&\, H_{a,b,c}
N_{a+1,b,c} \cr
E^{(2)}_{a,b,c}\, & = &\, N_{a,b,c} V_{a,b,c+1}\, &  - &\,
V_{a,b,c} N_{a,b+1,c} \cr
E^{(3)}_{a+1,b+1,c+1}\, & = &\, N_{a+1,b+1,c} D_{a+1,b+1,c+1}\, & -&\,
D_{a+1,b+1,c} N_{a,b,c-1}.\cr}}
These can be read from the triangles in figures 6A, 6B and 6F,
respectively. The triangles can be thought of as describing the Yukawa
interactions in \lagrferm.

These functions satisfy the relation
\eqn\constrejone{\sum_{i}\sum_{a,b,c} E^{(i)}_{a,b,c} J^{(i)}_{a,b,c}\,=\,
0.}

\bigskip
\centerline{\vbox{\hsize=5in\tenpoint
\centerline{\psfig{figure=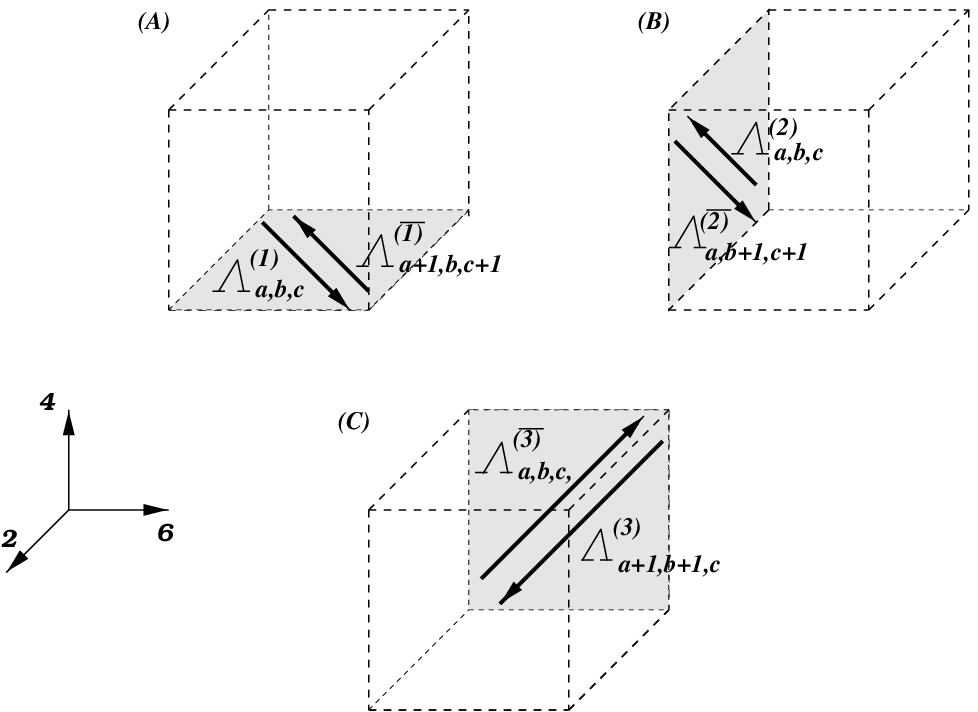}}
\vglue .4in
{\ninerm Fig. 5.  The arrows associated to (0,2) Fermi multiplets and
their conjugates. The fields represented correspond  to
$\Lambda^{(1)}_{a,b,c}$, $\Lambda^{({\bar 1})}_{a+1,b,c+1}$ (figure A)
$\Lambda^{(2)}_{a,b,c}$, $\Lambda^{({\bar 2})}_{a,b+1,c+1}$ (figure B),
and
$\Lambda^{(3)}_{a+1,b+1,c}$, $\Lambda^{({\bar 3})}_{a,b,c}$ (figure C).}}
}

Observe the analogy between both types of functions and in the
interactions they give rise to. This is the main motivation to introduce
the same diagrammatic representation for both. Actually, at the level of
component fields, the interactions are very similar. It is only the
choice of a specific chiral superfield to appear in the functions $E$ of the
Fermi multiplets (in this case, the fields $N$) that introduces the
difference in the origin of the terms, as some arising from the
superpotential, and others from the Lagrangian $L_F$.

We stress that the same theory could be rewritten in $(0,2)$ superspace
with a different choice for this special field. In that case, there would
be some re-shuffling of interactions, and terms which originate in the
$(0,2)$
superpotential for one choice can appear from $L_F$ for another choice. In
the case we have studied, it was natural to take the fields $N$ as
special, since they were in the adjoint, and the $(0,2)$ superspace
Lagrangian obtained with this choice can be further written in $(2,2)$
superspace. For more general $(0,2)$ theories, to be studied in next
section, there is no canonical choice of special field.

\vskip 1truecm
\subsec{General Rules to Obtain the Two-Dimensional Field Theory}

In this section we consider a more general model, with a unit cell
formed by $k\times k'\times k''$ boxes.
In each box we can place an arbitrary number of D4 branes, denoted
$n_{a,b,c}$ (with $a=1,2,\dots,k$, $b=1,2,\dots,k'$ and $c=1,2,\dots,k''$)
for the box labeled $(a,b,c)$. Indices $a,b$ and $c$ correspond
to positions along the directions 6,4 and 2 respectively.
Notice that as we define the theory on the
torus ${\bf T}^3$ the indices
$a,b,c$ are defined modulo $k,k',k''$ respectively.

The rules we determined in the previous section have a natural
generalization to this more general case. In the following we state these
general rules, and then present some arguments supporting it.

The gauge group and matter content are specified by $k,k',k''$ and the set
of numbers
$\{n_{a,b,c}\}$. The gauge group associated to this configuration is given by
$\prod_{a,b,c} U(n_{a,b,c})$.

\medskip

The matter content of the model consists of the following
$(0,2)$ chiral multiplets.
{\parindent=1cm
\item{$\bullet$} The `horizontal' multiplet $H_{a,b,c}$,  which transforms
in the
bi-fundamental representation  $(\fund,\antifund)$
of $U(n_{a,b,c}) \times$ $U(n_{a+1,b,c})$. It is represented by an arrow
going from the box $(a,b,c)$ to the box $(a+1,b,c)$.

\item{$\bullet$} The `vertical' multiplet $V_{a,b,c}$,  transforming in
the $(\fund,\antifund)$ of $U(n_{a,b,c}) \times$ $U(n_{a,b+1,c})$. It
corresponds to an arrow going from the box $(a,b,c)$ to the box
$(a,b+1,c)$.

\item{$\bullet$} The `normal' multiplet $N_{a,b,c}$, in the
$(\fund,\antifund)$ of $U(n_{a,b,c}) \times$ $U(n_{a,b,c+1})$. Its
arrow, from the box $(a,b,c)$ to the box $(a,b,c+1)$, is normal to those
of $H$ and $V$.

\item{$\bullet$} The `diagonal' multiplet $D_{a,b,c}$, which transforms in
the $(\fund,\antifund)$ of $U(n_{a,b,c}) \times$ $U(n_{a-1,b-1,c-1})$.
Its arrow goes from the box $(a,b,c)$ to the box $(a-1,b-1,c-1)$.
}

The arrows corresponding to these chiral fields are depicted in
Figure ~4.

\medskip

The theory also contains the following $(0,2)$ Fermi multiplets.

{\parindent=1cm
\item{$\bullet$} The conjugate fields $\Lambda^{(1)}_{a,b,c}$,
$\Lambda^{({\bar 1})}_{a+1,b,c+1}$, which transform respectively in the
$(\fund,\antifund)$ and $(\antifund,\fund)$ representations of
$U(n_{a,b,c})$$\times U(n_{a+1,b,c+1})$. They are represented by arrows
going from the box $(a,b,c)$ to the box $(a+1,b,c+1)$, and vice versa.

\item{$\bullet$} The conjugate fields $\Lambda^{(2)}_{a,b,c}$,
$\Lambda^{(\bar 2)}_{a,b+1,c+1}$, transforming respectively in the
$(\fund,\antifund)$ and $(\antifund,\fund)$ of $U(n_{a,b,c})$$ \times
U(n_{a,b+1,c+1})$. They are represented by arrows from the box $(a,b,c)$
to the box $(a,b+1,c+1)$ and vice versa.

\item{$\bullet$} The conjugate fields $\Lambda^{(3)}_{a,b,c}$,
$\Lambda^{(\bar 3)}_{a-1,b-1,c}$, transforming respectively in the
$(\fund,\antifund)$ and $(\antifund,\fund)$ of $U(n_{a,b,c})$$ \times
U(n_{a-1,b-1,c})$. They are represented by arrows from the box $(a,b,c)$
to the box $(a-1,b-1,c)$ and vice versa.
}

The arrows representing the Fermi supermultiplets and their conjugates are
shown in Figure ~5. The complete spectrum of matter multiplets is shown in
Table~2.

\medskip

\midinsert{
$$
{\sevenrm
\vbox{\offinterlineskip
\def\strut{\vrule height 3.25ex  width 0pt depth 2ex}
\def\hline{\noalign{\hrule}}
\halign{\vrule#\hfil\strut&\quad#\hfil\quad\vrule&&\quad\enspace
\hfil#\hfil\quad\vrule\cr
\noalign{\hrule}
&$\hfil{\bf Field}$& ${\bf Representation}$\cr
\noalign{\hrule}
&$H_{a,b,c}$& $(\fund_{a,b,c}, \antifund_{a+1,b,c})$\cr
\hline
& $V_{a,b,c}$& $(\fund_{a,b,c}, \antifund_{a,b+1,c})$\cr
\hline
& $N_{a,b,c}$& $(\fund_{a,b,c}, \antifund_{a,b,c+1})$ \cr
\hline
&$D_{a,b,c}$& $(\fund_{a,b,c}, \antifund_{a-1,b-1,c-1})$\cr
\hline
& $\Lambda^{(1)}_{a,b,c}$& $(\fund_{a,b,c}, \antifund_{a+1,b,c+1})$\cr
& $\Lambda^{({\bar 1})}_{a+1,b,c+1}$& $(\fund_{a+1,b,c+1},
\antifund_{a,b,c})$\cr
\hline
& $\Lambda^{(2)}_{a,b,c}$& $(\fund_{a,b,c}, \antifund_{a,b+1,c+1})$\cr
& $\Lambda^{({\bar 2})}_{a,b+1,c+1}$& $(\fund_{a,b+1,c+1},
\antifund_{a,b,c})$\cr
\hline
& $\Lambda^{(3)}_{a,b,c}$& $(\fund_{a,b,c}, \antifund_{a-1,b-1,c})$\cr
& $\Lambda^{({\bar 3})}_{a-1,b-1,c}$& $(\fund_{a-1,b-1,c},
\antifund_{a,b,c})$\cr
\hline
}}}
$$ }
{\ninerm
{Table 2:  Spectrum of chiral matter obtained from the general rules. The
table shows the representation of the gauge group in which each field is
transforming. We have listed the Fermi multiplets along with their
conjugates.}}
\endinsert

We again stress that there are only three independent Fermi
multiplets, the others being merely conjugates of them. However there is
no canonical choice of `fundamental' and `conjugate' fields. This
ambiguity is actually related to the choice of which interactions arise
from the $(0,2)$ superpotential from $L_J$ and which from the Lagrangian $L_F$. This
splitting of the interactions is determined by the choice of a `special'
chiral multiplet, which will appear in all the functions $E$ associated to
the Fermi multiplets. The rules below can provide the $(0,2)$ Lagrangian
for any choice, but for the ease of explanation we will write the
explicit equations only for the case the fields $D_{a,b,c}$ are taken as
the special ones. In order to clarify this issue and to complete our
set of rules, let us turn to the interactions in the theory.

They are given by closed oriented triangles of arrows in the brane
diagram. The possible triangles are depicted in Figure~6. Once a special
chiral multiplet is chosen, the interactions
are arranged in two sets, those arising from the $(0,2)$ superpotential,
and those from the $E$ functions associated to the Fermi multiplets. Since
the special field appear in all the $E$ functions, the triangles in which
it appears give interactions coming from $L_F$; below we give rules to
compute the corresponding functions $E$. The triangles where the
special field does not appear provide the $(0,2)$ superpotential. We will
give rules to compute the corresponding functions $J$. Since the
superpotential \seis\ is holomorphic, we can read from the triangles whose
fields $\Lambda$ are the Fermi multiplets and which are the conjugates.

\bigskip
\centerline{\vbox{\hsize=5in\tenpoint
\centerline{\psfig{figure=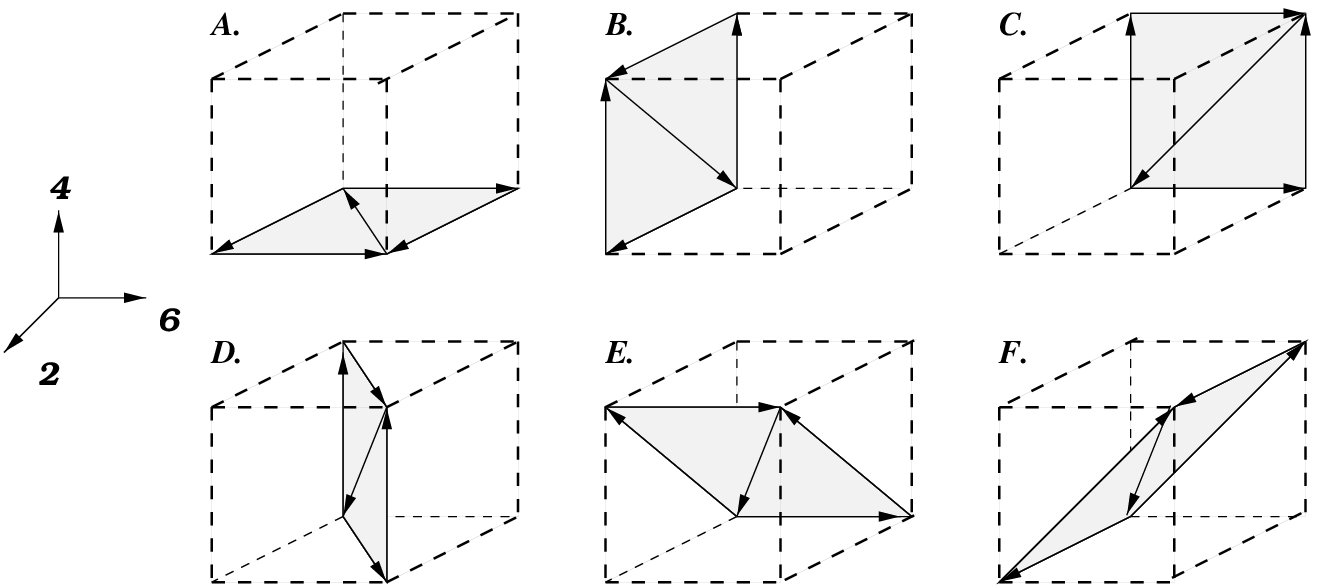}}
\vglue .4in
{\ninerm Fig. 6.  Prescriptions to compute the interactions in the (0,2)
theory. The twelve triangles depicted encode the definition of the
functions $E$ and $J$ corresponding to the Fermi multiplets.}}}


Let us choose the fields $D_{a,b,c}$ to be the special fields. This means
that, in figure ~6, the triangles 6A, 6B and 6C provide the $(0,2)$
superpotential. It also means that the Fermi superfields are
$\Lambda^{({\bar 1})}$, $\Lambda^{({\bar 2})}$ and $\Lambda^{(3)}$,
whereas  $\Lambda^{(1)}$, $\Lambda^{(2)}$ and $\Lambda^{({\bar 3})}$ are
their conjugates. We can read the corresponding functions $J$ from the
triangles.

\eqn\jfuncttwo{\matrix{
\Lambda^{({\bar 1})}_{a+1,b,c+1} \; & \to \;\; J^{({\bar
1})}_{a+1,b,c+1} & =\, H_{a,b,c} N_{a+1,b,c} \, -\, N_{a,b,c} H_{a,b,c+1}
\cr
\Lambda^{({\bar 2})}_{a,b+1,c+1} \; & \to \;\; J^{({\bar
2})}_{a,b+1,c+1} & =\, N_{a,b,c}V_{a,b,c+1} \, - \,
V_{a,b,c} N_{a,b+1,c}  \cr
\Lambda^{(3)}_{a+1,b+1,c}\; & \to \;\;
J^{(1)}_{a+1,b+1,c} & =\, V_{a,b,c}H_{a,b+1,c} \,
-\,H_{a,b,c}V_{a+1,b,c}.\cr
}}

The remaining triangles, 6D, 6E and 6F, involve the fields $D$,
and so give interactions arising from $L_F$. The triangles provide the $E$
functions corresponding to the Fermi fields:

\eqn\efuncttwo{\matrix{
\Lambda^{({\bar 1})}_{a+1,b,c+1} \;& \to \;\; E^{({\bar
1})}_{a+1,b,c+1} & =\, D_{a+1,b,c+1} V_{a,b-1,c} \, -\,
V_{a+1,b,c+1} D_{a+1,b+1,c+1} \cr
\Lambda^{({\bar 2})}_{a,b+1,c+1}\; & \to \;\; E^{({\bar 2})}_{a,b+1,c+1}
& =\, D_{a,b+1,c+1}H_{a-1,b,c} \, -\, H_{a,b+1,c+1} D_{a+1,b+1,c+1} \cr
\Lambda^{(3)}_{a+1,b+1,c}\; & \to \;\; E^{(3)}_{a+1,b+1,c} & = \,
D_{a+1,b+1,c} N_{a,b,c-1} \, -\, N_{a+1,b+1,c} D_{a+1,b+1,c+1}.  \cr
}}

One can check that they verify the relation
\eqn\constrejtwo{\sum_i \sum_{a,b,c} \big( J^{({\bar 1})}_{a,b,c}
E^{({\bar 1})}_{a,b,c} + J^{({\bar 2})}_{a,b,c} E^{({\bar 2})}_{a,b,c} +
J^{(3)}_{a,b,c} E^{(3)}_{a,b,c} \big) \; =\; 0.}

Since the functions $J$ and $E$ specify the interactions, we
have completed the characterization of the $(0,2)$ field theory. It is a
straightforward matter to extract the interactions in components by using
the relevant formulae in Section~2. The characterization of the field
theory parameters in terms of the parameters in the brane configuration is
carried out in appendix I.

\medskip

This set of rules is the natural generalization of those determined for
the particular case of $(2,2)$ theories in Section~3.2. A first good
property of these rules is that whenever one of the chiral multiplets
appears in adjoint representations, the spectrum and interactions
will have enhanced $(2,2)$ supersymmetry. The simplest way of showing it
is to take that chiral multiplet as the `special' one. Then all
interactions arising from functions $E$ become gauge interactions in the
$(2,2)$ theory, and the couplings form the functions $J$ will correspond
to the $(2,2)$ superpotential. We will come back to this and related
points in Section~4.

Another check of our set of rules comes from the study of
Higgs breakings in the classical theory. It is clear in the brane
configuration that there are
some brane movements which can be performed. Consider for instance the
case with the same number of D4 branes in all boxes. Then the NS branes
can be moved along 7, the NS$'$ branes can be moved along 5, and the
NS$''$ branes can be moved along 3. As we argue in the appendix, these
movements correspond to changing the Fayet-Illiopoulos parameters in the
field theory. This, on the other hand, implies that some field in the
bi-fundamental gets a vev in order to make the D-term vanish and maintain
unbroken supersymmetry. This triggers gauge symmetry breaking, and as can
be checked in the field theory we have proposed, typically pairs of group
factors break to the diagonal subgroup. This is precisely the phenomenon
observed in the brane picture, and provides support to our identification
of the field theory. The analysis is similar to that in \zaffa\ and
we will not repeat the exercise here.

Another type of Higgs breaking corresponds to recombining D4 branes in
different boxes until they complete a set that can separate from the grid
of NS fivebranes. For instance, if all boxes have equal number of
D4 branes, there is a Higgs branch in which the recombined D4 branes wrap the
three-torus completely, and move freely in 35789. Other branches are
discussed in analogy with \hanur. Of course, our discussion has been in
terms of classical language. In two dimensions there is no moduli space of
vacua, and our `Higgs branches' should be understood as the target spaces
of the two-dimensional field theory interpreted as a sigma model. The
nature of this target space will be further clarified in Section~5.

A related point that usually arises in the context of $(2,2)$ and 
$(0,2)$ theories is the
dependence of the Higgs branch with the FI parameters. In particular, we
could ask whether our models will present phase transitions of the kind 
analyzed in \witten, establising some kind of Calabi-Yau/Landau-Ginzburg
correspondence. The field theories are rather complicated, and this study
is far from straightforward. In Section~5 we will clarify this issue by  
showing the relation of our field theories to those arising from D-branes
at singularities. This allows us to apply several results about the phase
structure of these linear sigma models \refs{\dgm,\mohri}.

Finally, let us mention that, since these theories are chiral, there are
potential gauge anomalies for the $U(1)$ factors. Actually, even for the
simplest choice of equal number of D4 branes in each box, the field theory
is seemingly anomalous. This issue has been studied in \mohri, -- in a  
T-dual picture of D1 branes at four-fold singularities, see section~5 --,
where the anomaly was seen to be cancelled through an interaction with bulk   
modes. In the brane box picture, we expect some kind of anomaly inflow   
mechanism playing an analogous role. We leave this very interesting point
for future research, and in the following will assume that such
mechanism is at work and renders the theory consistent.

\vskip 1truecm

\subsec{Examples}

\midinsert{
$$
{\sevenrm
\vbox{\offinterlineskip
\def\strut{\vrule height 3.25ex  width 0pt depth 2ex}
\def\hline{\noalign{\hrule}}
\halign{\vrule#\hfil\strut&\quad#\hfil\quad\vrule&&\quad\enspace
\hfil#\hfil\quad\vrule\cr
\noalign{\hrule}
&$\hfil {\bf Field}$& ${\bf Representation}$& ${\bf Field}$\cr
\noalign{\hrule}
\hline
&$H_{1,1,1}$& $(\fund,\antifund,1,1,1,1,1,1)$& $\bar{H}_{2,1,1}$\cr
\hline
&$H_{1,2,1}$& $(1,1,\fund,\antifund,1,1,1,1)$& $\bar{H}_{2,2,1}$\cr
\hline
&$H_{1,1,2}$& $(1,1,1,1,\fund,\antifund,1,1)$& $\bar{H}_{2,1,2}$\cr
\hline
&$H_{1,2,2}$& $(1,1,1,1,1,1,\fund,\antifund)$& $\bar{H}_{2,2,2}$\cr
\hline
\hline
&$V_{1,1,1}$& $(\fund,1,\antifund,1,1,1,1,1)$& $\bar{V}_{1,2,1}$\cr
\hline
&$V_{2,1,1}$& $(1,\fund,1,\antifund,1,1,1,1)$& $\bar{V}_{2,2,1}$\cr
\hline
&$V_{1,1,2}$& $(1,1,1,1,\fund,1,\antifund,1)$& $\bar{V}_{1,2,2}$\cr
\hline
&$V_{2,1,2}$& $(1,1,1,1,1,\fund,1,\antifund)$& $\bar{V}_{2,2,2}$\cr
\hline
\hline
&$N_{1,1,1}$& $(\fund,1,1,1,\antifund,1,1,1)$& $\bar{N}_{1,1,2}$\cr
\hline
&$N_{2,1,1}$& $(1,\fund,1,1,1,\antifund,1,1)$& $\bar{N}_{2,1,2}$\cr
\hline
&$N_{1,2,1}$& $(1,1,\fund,1,1,1,\antifund,1)$& $\bar{N}_{1,2,2}$\cr
\hline
&$N_{2,2,1}$& $(1,1,1,\fund,1,1,1,\antifund)$& $\bar{N}_{2,2,2}$\cr
\hline
\hline
&$D_{1,1,1}$& $(\fund,1,1,1,1,1,1,\antifund)$& $\bar{D}_{2,2,2}$\cr
\hline
&$D_{2,1,1}$& $(1,\fund,1,1,1,1,\antifund,1)$& $\bar{D}_{1,2,2}$\cr
\hline
&$D_{1,2,1}$& $(1,1,\fund,1,1,\antifund,1,1)$& $\bar{D}_{2,1,2}$\cr
\hline
&$D_{2,2,1}$& $(1,1,1,\fund,\antifund,1,1,1)$& $\bar{D}_{1,1,2}$\cr
\hline
\hline
}}}
$$ }
{\ninerm
Table 3: Chiral multiplets for the $2\times 2\times 2$ box model. The
ordering of the $U(n)$ factors in the second column is
$U(n)_{1,1,1}$$\times
U(n)_{2,1,1}$$\times U(n)_{1,2,1}$$\times U(n)_{2,2,1}$$\times
U(n)_{1,1,2}\times$$U(n)_{2,1,2}$$\times U(n)_{1,2,2}$$\times
U(n)_{2,2,2}$.}
\endinsert

To see how the above set of rules works we consider an specific example with
eight different three-dimensional boxes. Here we have two NS branes,
two NS' branes and two NS'' branes and an equal number $n$ of D4 branes in
each box. The theory is defined on ${\bf T}^3$ which arises from the grid
identified by shifts by two boxes in each of the directions $246$. Thus
the $2\times 2\times 2$ unit cell consist of eight
boxes, so the gauge group is $U(n)^8$. Even though the theory is $(0,2)$,
the matter content is vector-like. The chiral multiplets in the model are
listed in Table~3 (to make it shorter we list half or the
fields and the conjugates of the other half)

In order to list the Fermi supermultiplets, we must make a choice of
`special' chiral superfield. For concreteness we will pick the fields $D$
as the special ones. The corresponding Fermi multiplets are given in
Table~4.

The interactions can be obtained by straightforward application of Eqs.
\jfuncttwo\ and  \efuncttwo.

\midinsert{
{\sevenrm
$$
\vbox{\offinterlineskip
\def\strut{\vrule height 3.25ex  width 0pt depth 2ex}
\def\hline{\noalign{\hrule}}
\halign{\vrule#\hfil\strut&\quad#\hfil\quad\vrule&&\quad\enspace
\hfil#\hfil\quad\vrule\cr
\noalign{\hrule}
&$\hfil {\bf Field}$& ${\bf Representation}$& ${\bf Field}$ & {\bf
Representation} \cr
\noalign{\hrule}
\hline

&$\Lambda^{({\bar 1})}_{1,1,1}$ & $(\fund,1,1,1,1,\antifund,1,1)$&
$\Lambda^{({\bar 1})}_{2,1,2}$ & {\rm Conj. rep.}\cr
\hline
&$\Lambda^{({\bar 1})}_{2,1,1}$ & $(1,\fund,1,1,\antifund,1,1,1)$&
$\Lambda^{({\bar 1})}_{1,1,2}$ & {\rm Conj. rep.}\cr
\hline
&$\Lambda^{({\bar 1})}_{1,2,1}$ & $(1,1,\fund,1,1,1,1,\antifund)$&
$\Lambda^{({\bar 1})}_{2,2,2}$ & {\rm Conj. rep.}\cr
\hline
&$\Lambda^{({\bar 1})}_{2,2,1}$ & $(1,1,1,\fund,1,1,\antifund,1)$&
$\Lambda^{({\bar 1})}_{1,2,2}$ & {\rm Conj. rep.}\cr
\hline
\hline
&$\Lambda^{({\bar 2})}_{1,1,1}$ & $(\fund,1,1,1,1,1,\antifund,1)$&
$\Lambda^{({\bar 2})}_{1,2,2}$ & {\rm Conj. rep.}\cr
\hline
&$\Lambda^{({\bar 2})}_{1,2,1}$ & $(1,1,\fund,1,\antifund,1,1,1)$&
$\Lambda^{({\bar 2})}_{1,1,2}$ & {\rm Conj. rep.}\cr
\hline
&$\Lambda^{({\bar 2})}_{2,1,1}$ & $(1,\fund,1,1,1,1,1,\antifund)$&
$\Lambda^{({\bar 2})}_{2,2,2}$ & {\rm Conj. rep.}\cr
\hline
&$\Lambda^{({\bar 2})}_{2,2,1}$ & $(1,1,1,\fund,1,\antifund,1,1)$&
$\Lambda^{({\bar 2})}_{2,1,2}$ & {\rm Conj. rep.}\cr
\hline
\hline
&$\Lambda^{(3)}_{1,1,1}$ & $(\fund,1,1,\antifund,1,1,1,1)$&
$\Lambda^{(3)}_{2,2,1}$ & {\rm Conj. rep.}\cr
\hline
&$\Lambda^{(3)}_{2,1,1}$ & $(1,\fund,\antifund,1,1,1,1,1)$&
$\Lambda^{(3)}_{1,2,1}$ & {\rm Conj. rep.}\cr
\hline
&$\Lambda^{(3)}_{1,1,2}$ & $(1,1,1,1,\fund,1,1,\antifund)$&
$\Lambda^{(3)}_{2,2,2}$ & {\rm Conj. rep.}\cr
\hline
&$\Lambda^{(3)}_{2,1,2}$ & $(1,1,1,1,1,\fund,\antifund,1)$&
$\Lambda^{(3)}_{1,2,2}$ & {\rm Conj. rep.}\cr
\hline
\hline
}}
$$ }}
{\ninerm
{Table 4: Fermi multiplets for the $2\times 2\times 2$ box model. The
fields $D$ have been chosen as the special chiral multiplets, and the
Fermi superfields are of type $\Lambda^{({\bar 1})}$, $\Lambda^{({\bar
2})}$ and $\Lambda^{(3)}$. Recall that the ordering of the groups in the
second column is $U(n)_{1,1,1}$$\times U(n)_{2,1,1}$$\times U(n)_{1,2,1}$
$\times U(n)_{2,2,1}$$\times U(n)_{1,1,2}\times$$U(n)_{2,1,2}$$\times
U(n)_{1,2,2}$$\times U(n)_{2,2,2}$.}}
\endinsert

\vskip 2truecm
\newsec{Models with Enhanced Supersymmetry}

\subsec{Non-chiral enhancement of supersymmetry}

In this section we discuss how to construct two-dimensional gauge theories
with enhanced supersymmetry using the brane boxes introduced above. We
begin by briefly mentioning the simpler case of non-chiral supersymmetry.
We have already considered such theories in section~3.2, where we studied
the dimensional reduction of four-dimensional $\NN=1$ models, {\it i.e.}
$(2,2)$ theories in two dimensions. It is easy to see that the pattern of
the spectrum of the $(2,2)$ field theories that can be constructed in our
setup is that of the dimensional reductions of
four-dimensional box models. Here by `pattern' we mean the structure of
the
bi-fundamental multiplets, the quiver of the theory, regardless of the
actual rank of the gauge factors. We stress this subtlety because, since
$(2,2)$ theories are non-chiral, there is more freedom in choosing the
numbers of D4 branes in the boxes. So some of our $(2,2)$ models would be
dimensional reduction of anomalous four-dimensional theories. The point
here has already been stressed in \hanhori.

This relation to four-dimensional models greatly facilitates the
construction of brane models with enhanced $(2,2)$, $(4,4)$ or $(8,8)$
supersymmetry. One basically takes the brane
construction of four-dimensional $\NN=1$, $\NN=2$ or $\NN=4$ theories in
\refs{\zaffa,\math}, replaces the D5 branes by D4 branes,
and adds one NS$''$ brane. Thus for example, $(4,4)$ theories can be
obtained by considering $k'=k''=1$, and arbitrary $k$. The unit cell is a
finite row of $k$ boxes, with trivial identifications of its sides. One
such brane model is shown in Figure~7, for the particular case of
$k=4$. The
identification of the spectrum of the theory and the interactions can be
done directly in four-dimensional language, using the rules in \zaffa.

Similarly, gauge theories with sixteen supercharges can be obtained by
considering a unit cell with a single box, with all sides identified. All
fields in the theory are in the adjoint representation, and the
brane configuration can be interpreted as D4 branes wrapped on a
three-torus.

\bigskip
\centerline{\vbox{\hsize=4in\tenpoint
\centerline{\psfig{figure=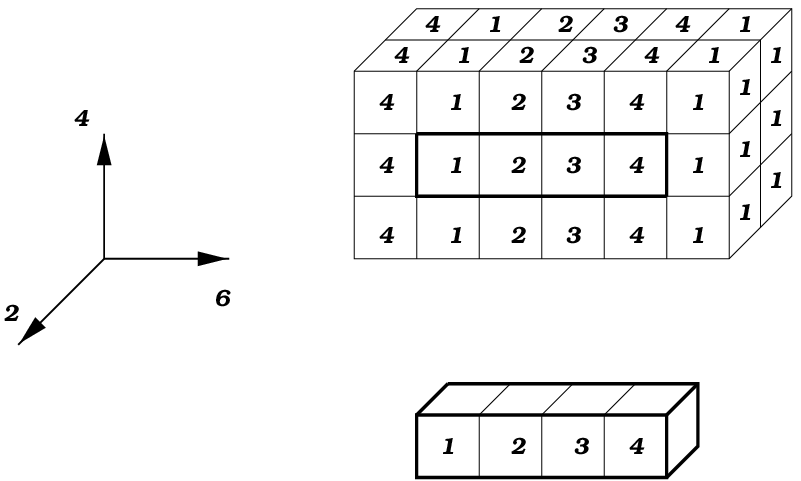}}
\vglue .2in
{\ninerm Fig. 7. The figure shows the brane box configuration which lead
to a gauge theory in two dimensions with (4,4) supersymmetry. The unit
cell consist of 4 boxes extending horizontally. Observe it has trivial
identifications of its faces in all three directions 642. Recall that the
numbers in the boxes are merely labels to distinguish them.}}}

\vskip 1truecm
\noindent
\subsec{Chiral Supersymmetry Enhancement}

{\it (0,4) Theories}

It is easy to construct brane box models with enhanced $(0,4)$
supersymmetry. The rule is that any model in which one kind of Fermi
multiplet appears in the adjoint has at least $(0,4)$ supersymmetry.
Instead of discussing it in general, let us present one such example which
illustrates the general features of these theories. The relevant facts
about the field theory will be mentioned as needed (the essentials about
$(0,4)$ supersymmetric theories can be extracted from \inst).

The configuration is the  $4 \times 1 \times 1$ box model depicted in
Figure~8. We label the boxes in the unit cell with {\it one} index $i$,
and denote by $n_i$ the number of D4 branes on each box. Notice that the
faces of the unit cell are identified up to shifts in several directions.
These non-trivial identifications of the sides of the unit cell can
be read from the picture.

\bigskip
\centerline{\vbox{\hsize=4in\tenpoint
\centerline{\psfig{figure=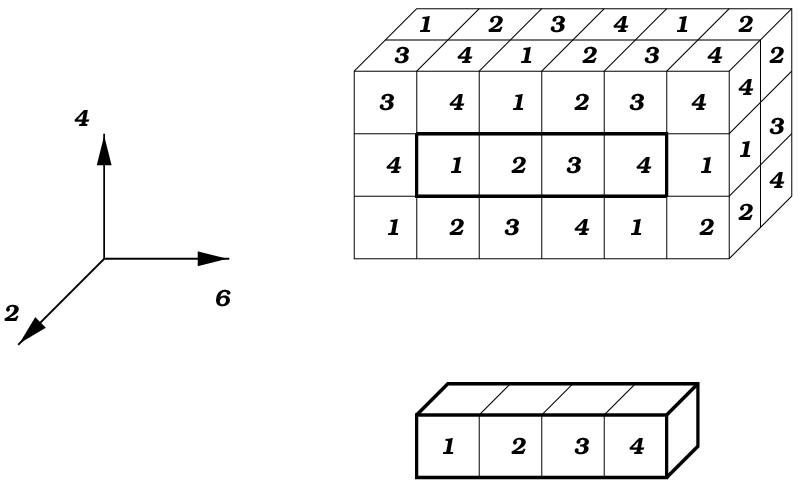}}
\vglue .2in
{\ninerm Fig. 8. The figure corresponds to a brane box configuration which
leads to a two-dimensional field theory with (0,4) supersymmetry. The unit
cell, also displayed, consists of 4 boxes extending along 6. Observe that
in defining the three-torus by identifying the faces of the unit cell, the
identifications are accompanied by shifts. These can be read from the
figure. The model is said to have non-trivial identifications.}}}

The gauge group is $\prod_i U(n_i)$. From our general rules we can read
the matter content, which we have collected in Table~5 (the Fermi
multiplets shown correspond to the choice of $D$ as special field). We
have also listed the $(0,4)$ multiplets these fields form.

For instance,
the fields $\Lambda^{(3)}$, which transform in the adjoint, become part of
the $(0,4)$ vector multiplet. These fermions, along with the $(0,2)$
gauginos, can be arranged in a four-dimensional vector, acted upon by the
$SO(4)_R$ R-symmetry group of the field theory. Notice that only the usual
$U(1)_R$ subgroup is visible in the brane configuration.

The fields $H_i$ and $V_{i+1}$ combine to
form one $(0,4)$ chiral multiplet (since they transform in conjugate
representations). The fields $N_i$ and $D_{i+2}$ form another $(0,4)$
chiral multiplet. The Fermi
multiplets $\Lambda^{({\bar 1})}$ and $\Lambda^{({\bar 2})}$ remain as
two $(0,4)$ Fermi multiplets, singlets under the R-symmetry.

\midinsert{
$$
{\sevenrm
\vbox{\offinterlineskip
\def\strut{\vrule height 3.25ex  width 0pt depth 2ex}
\def\hline{\noalign{\hrule}}
\halign{\vrule#\hfil\strut&\quad#\hfil\quad\vrule&&\quad\enspace
\hfil#\hfil\quad\vrule\cr
\noalign{\hrule}
&$\hfil{\bf (0,2) \ Field}$& ${\bf Representation}$ & ${\bf (0,4) \ Field}$\cr
\noalign{\hrule}
&$H_i$&$(\fund_i,\antifund_{i+1})$&  ${\rm Chiral}$ \cr
&$V_{i+1}$& $(\fund_{i+1},\antifund_{i})$& \cr
\hline
&$N_i$&$(\fund_i,\antifund_{i+2})$&  ${\rm Chiral}$\cr
&$D_{i+2}$& $(\fund_{i+2},\fund_{i})$& \cr
\hline
&$\Lambda^{({\bar 1})}_i$&$(\fund_i,\antifund_{i+1})$ & ${\rm Fermi}$\cr
\hline
&$\Lambda^{({\bar 2})}_i$&$(\fund_i,\antifund_{i-1})$&${\rm Fermi}$\cr
\hline
&$\Lambda^{(3)}_i$&$(\fund_i,\antifund_{i})$&  ${\rm Vector}$\cr
\hline
}}}
$$ }
{\ninerm
{Table 5: The table summarizes the matter content corresponding to the
brane box configuration of Figure~8. The only field in the adjoint
representation
corresponds to a Fermi field. The rest of fields are charged as
bifundamentals. The model has (0,4) supersymmetry, and in the last column
we show how the $(0,2)$ fields combine in $(0,4)$ multiplets. }}
\endinsert

Concerning the interactions, they also respect $(0,4)$ supersymmetry. For
instance, for the Fermi
multiplet in the adjoint $\Lambda^{(3)}_i$ we have

\eqn\ezerofour{\matrix{
J^{(3)}_i & = & H_i V_{i+1} - V_i H_{i-1}\cr
E^{(3)}_i & = & D_i N_{i-2} - N_i D_{i+2}.}}

It is easy to see that, when expressed in components, the Yukawa couplings
from these terms are the `gauge' interactions of the $(0,4)$
chiral multiplet with the additional gaugino $\lambda^{(3)}_i$ in the
$(0,4)$ vector multiplet.

Let us also discuss the interactions of the remaining Fermi multiplets.
The main novelty is that the two kinds of couplings of a
Fermi multiplet to the chiral multiplets, given by the functions $E$ and
$J$, become related by supersymmetry. For instance, for the multiplet
$\Lambda^{({\bar 1})}_i$, we have

\eqn\zerofourprime{\matrix{
J^{({\bar 1})}_i & = & N_{i+1} H_{i+3} - H_{i+1} N_{i+2} \cr
E^{({\bar 1})}_i & = & D_i V_{i-2} - V_i D_{i-1}.
}}
The corresponding terms are the interaction of the Fermi multiplet
$\Lambda^{({\bar 1})}$ with
the $(0,4)$ multiplets formed by the pairs $(H_i,V_{i+1})$, and $(N_i,
D_{i+2})$.

Analogously, the functions associated to the Fermi multiplet
$\Lambda^{({\bar 2})}_i$,

\eqn\zerofoursecond{\matrix{
J^{({\bar 2})}_i & = & V_{i-1} N_{i-2} - N_{i-1} V_{i+1} \cr
E^{({\bar 2})}_i & = & D_i H_{i-2} - H_i D_{i+1},
}}
contain the interactions of $\Lambda^{(2)}$ with the $(0,4)$ chiral
multiplets.

\medskip

This pattern holds in general. Any time there is one kind of Fermi
multiplet in adjoint representations, the whole spectrum fits nicely into
$(0,4)$ multiplets. It can also be checked that the interactions respect
this supersymmetry. In order to avoid non-chiral supersymmetry enhancement
there should not be any chiral multiplet in the adjoint, otherwise the
supersymmetry would be $(4,4)$. There are many other models that can be
constructed in a similar way, but we will not attempt a general
classification. The supersymmetry enhancement of these models will receive
a nice geometrical interpretation in section~5.4.

\vskip 1truecm
\noindent
{\it (0,6) Theories}

It is also possible to get two-dimensional theories with six chiral
supercharges. The appropriate way of obtaining them is constructing
brane configurations which have two Fermi multiplets transforming in the
adjoint. If the supersymmetry enhancement is to be chiral, one must also
make sure that there are no chiral multiplets in adjoint representations.

In the following we present an example of one such model. The
configuration is a $4 \times 1 \times 1$ box model, which we show in
Figure~9. We label the boxes in the unit cell by an index $i$, as before.

The gauge group is $\prod_i U(n_i)$. The matter content is summarized in
Table~6, where we also show how the $(0,2)$ fields combine to form
$(0,6)$ multiplets.

The R-symmetry group is $SO(6)_R$. All four $(0,2)$ chiral multiplets
combine into a single $(0,6)$ chiral multiplet. The four complex fields
transform in the fundamental representation of $SU(4)_R\approx SO(6)_R$.

Also, $\Lambda^{({\bar 2})}$ and
$\Lambda^{(3)}$, which transform in the adjoint, become part of the vector
multiplet. Along with the gauginos in the $(0,2)$ multiplet, they can be
arranged in a six dimensional real vector, acted upon by $SO(6)_R$

The field $\Lambda^{({\bar 1})}$ remains a Fermi multiplet.
Again, the couplings also respect the higher supersymmetry. For instance
the interactions of $\Lambda^{({\bar 2})}$ and $\Lambda^{(3)}$ become
have the appropriate structure to be `gauge' interactions.

\medskip

It is easy to construct the most general brane box model with this
structure. It is given by a generalization of the model in Figure~9, by
considering a unit cell with an arbitrary number $k$ of boxes. The
spectrum is also given by Table~6 by simply increasing the range of
variation of $i$, $i=1,\ldots,k$.

\bigskip
\centerline{\vbox{\hsize=4in\tenpoint
\centerline{\psfig{figure=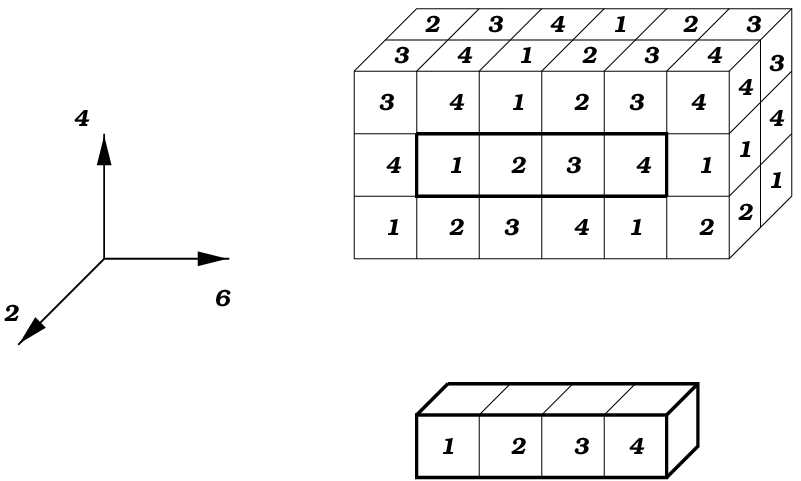}}
\vglue .2in
{\ninerm Fig. 9. The figure shows a $4 \times 1 \times 1$ box model which
leads to the construction of two-dimensional gauge theories with (0,6)
supersymmetry. The unit cell consists of 4 boxes, and its faces are
identified in a non-trivial way.}}}

\midinsert{
{\sevenrm
$$
\vbox{\offinterlineskip
\def\strut{\vrule height 3.25ex  width 0pt depth 2ex}
\def\hline{\noalign{\hrule}}
\halign{\vrule#\hfil\strut&\quad#\hfil\quad\vrule&&\quad\enspace
\hfil#\hfil\quad\vrule\cr
\noalign{\hrule}
&$\hfil{\bf (0,2) \ Field}$& ${\bf Representation}$ & ${\bf (0,6) \ Field}$\cr
\noalign{\hrule}
&$H_i$&$(\fund_i,\antifund_{i+1})$&  \cr
&$V_{i+1}$& $(\fund_{i+1},\antifund_{i})$& ${\rm Chiral}$ \cr
&$N_i$&$(\fund_i,\antifund_{i+1})$&  \cr
&$D_{i+1}$& $(\fund_{i+1},\antifund_{i})$& \cr
\hline
&$\Lambda^{({\bar 1})}_i$&$(\fund_i,\antifund_{i-2})$&${\rm Fermi}$\cr
\hline
&$\Lambda^{({\bar 2})}_i$&$(\fund_i,\antifund_{i})$&  ${\rm Vector}$\cr
&$\Lambda^{(3)}_i$&$(\fund_i,\antifund_{i})$&\cr
\hline
}}
$$ }}
{\ninerm
{Table 6: The table gives the matter content corresponding to the
brane box configuration of Fig. 9. There are two Fermi multiplets in the
adjoint representation. The rest of fields are charged as
bifundamentals. It corresponds to a model with (0,8) supersymmetry. }}
\endinsert

\vskip 1truecm
\noindent
{\it (0,8) Model}

Finally, let us construct the only $(0,8)$ model which can be realized in
our setup. The brane box configuration is given in Figure~10. The unit
cell contains two boxes, and has non-trivial identifications of its faces.
The gauge group is $U(n)^2$, and we easily check that all $(0,2)$ Fermi
multiplets are in the adjoint representation. They become part of the
$(0,8)$ gauge multiplet. The R-symmetry group $SO(8)_R$ acts on the
eight-dimensional real vector formed by these fermions and the $(0,2)$
gauginos. Also, all $(0,2)$ chiral multiplets transform in
the bifundamental $(\fund_1,\antifund_2)$ or its conjugate. The
chiral multiplets will fill $(0,8)$ chiral multiplets.
In this case, all the $(0,2)$ interactions are `gauge' interactions from
the $(0,8)$ point of view. The basic features of $(0,8)$ theories have been
determined {\it e.g.} in \bss.

\bigskip
\centerline{\vbox{\hsize=4in\tenpoint
\centerline{\psfig{figure=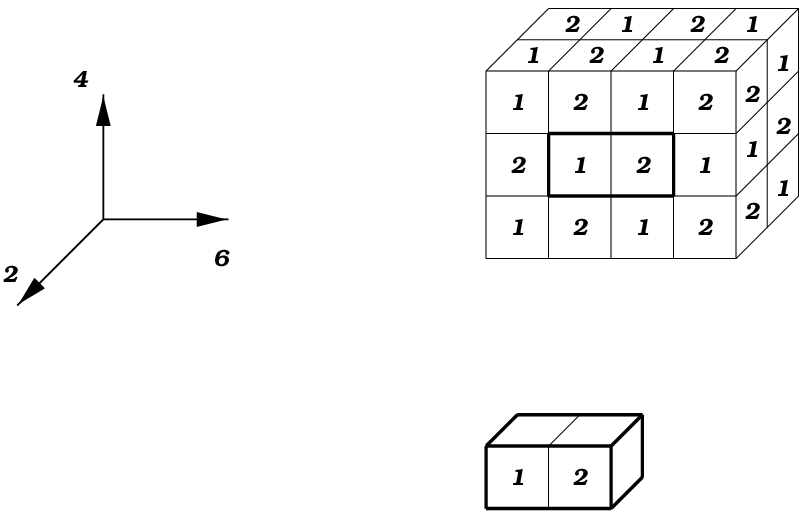}}
\vglue .2in
{\ninerm Fig. 10. The figure shows a $2 \times 1 \times 1$ box model
giving a $(0,8)$ field theory. The unit cell contains two boxes, and its
faces are identified in a non-trivial way.}}}

\medskip

The basic lesson we can learn from these further examples is that
enhancement of supersymmetry is easily obtained in our setup. Non-chiral
supersymmetries arise when one or several chiral multiplets transform in
the adjoint of the gauge group. Enhancement of chiral supersymmetries
appears when Fermi multiplets transform in the adjoint (and chiral
multiplets do not). Notice that the chiral enhancement of supersymmetry is
not manifest from the brane box point of view, and has to be checked by
direct computation of the spectrum and interactions. In section~5.4 we will
discuss how it becomes manifest in a T-dual configuration, where the field
theory is realized in the world-volume of D1 branes at
four-fold singularities. To establish such T-duality is the purpose of the
following section.

\vskip 2truecm


\newsec{The Interpretation of the Linear Sigma Model}

In the previous sections we have introduced a large family of
two-dimensional $(0,2)$ gauge theories. Since $(2,2)$ and $(0,2)$ theories
have been traditionally used as world-sheet descriptions of string
theories propagating on some target space, it is a natural
question
whether the (classical) Higgs branch or our models has any geometrical
interpretation
of the kind. In this section we are to show that it describes the
dynamics of a type IIB D1 brane on a ${\bf C}^4/\Gamma$ singularity, with
$\Gamma$ an abelian subgroup of $SU(4)$. The main tool for reaching this
conclusion will be a T-duality performed on the brane box model along the
directions 246.

\vskip 1truecm
\subsec{T-duality Along 246}

In this section we perform a T-duality on the brane box models along the
directions 246. The main tool will be the well known T-duality relation
between a set of $n$ parallel NS fivebranes and $n$ Kaluza-Klein
monopoles. The discussion in this subsection parallels that in \hanur.

Let us start with the simplest case of a brane box model formed by a unit
cell of $k \times k' \times k''$ boxes, with trivial identifications of
faces. In this case the T-duality along the directions 246 is
particularly
easy. We start with $k$ NS branes along 012345, $k'$ NS$'$ branes along
012367, and $k''$ NS$''$ branes along 014567. The T-duality along 2
transforms the NS$''$ branes into $k''$ Kaluza-Klein monopoles. These will
be described by a multi-center Taub-NUT metric, with non-trivial geometry
on the directions 2$'$,3,8,9, with 2$'$ denoting the coordinate dual to 2.
Notice that, since the 3,8,9 coordinates of the initial NS$''$ branes
coincided, so do the coordinates of the corresponding $k''$ centers of the
Taub-NUT metric, so that it contains singularities of type
$A_{k''-1}$.

Similarly, the T-duality along 4 transforms the $k'$ NS$'$ branes into
$k'$
Kaluza-Klein monopoles with world-volume along 012367, and represented by
a nontrivial geometry on 4$'$,5,8,9. Again, since the centers of the
Kaluza-Klein monopoles coincide, such geometry will contain singularities
of type $A_{k'-1}$. Finally, the T-duality along 6 turns the $k$ NS
branes into $k$ Kaluza-Klein monopoles. Their world-volume spans 012345,
and they are represented by a non-trivial geometry along 6$'$,7,8,9. Since
again all the centers coincide, there will be $A_{k-1}$ singularities.

Thus, the final T-dual of the grid of NS, NS$'$ and NS$''$ branes is
type IIB string theory with a complicated geometry in the directions
2$'$,3,4$'$,5,6$'$,7,8,9. One can
think of it roughly as some `superposition' of the Kaluza-Klein monopoles
we have described. Even without a quantitative knowledge of such metric,
we can describe the relevant features for our purposes. One such
feature is that the number of unbroken supersymmetries constrains
the manifold to be a Calabi-Yau four-fold. Also, from our remarks
above we know the existence of certain (complex) surfaces of
singularities of type $A_{k-1}$, $A_{k'-1}$ and $A_{k''-1}$ singularities.
If we introduce complex coordinates $w_1=\exp (x^7+ix^{6'})$
,$w_2=\exp(x^5+ix^{4'})$, $w_3=\exp(x^3+ix^{2'})$, and $w_4=x^9+ix^8$, the
surface of $A_{k-1}$ singularities is defined roughly by $w_1=w_4=0$,
the surface of $A_{k'-1}$ singularities is defined by $w_2=w_4=0$, and the
surface of $A_{k''-1}$ singularities is given by $w_3=w_4=0$. At the
origin $w_1$$=w_2$$=w_3$$=w_4=0$ all surfaces meet and the singularity is
worse. It can be described as a quotient singularity of type
$\IC^4/\Gamma$, with $\Gamma=\IZ_k\times \IZ_{k'}\times \IZ_{k''}$.
This
discrete group is generated by elements $\theta$, $\omega$, $\eta$, whose
action on $(z_1,z_2,z_3,z_4)\in \IC^4$ is as follows:

\eqn\transf{\matrix{
\theta: & (z_1,z_2,z_3,z_4) & \to  & (e^{2\pi i/k}z_1,z_2,z_3,e^{-2\pi i/k}z_4) \cr
\omega: & (z_1,z_2,z_3,z_4) & \to  & (z_1,e^{2\pi i/k'}z_2,z_3,e^{-2\pi i/k'}z_4) \cr
\eta: & (z_1,z_2,z_3,z_4) & \to & (z_1,z_2,e^{2\pi i/k''}z_3,e^{-2\pi i/k''}z_4).  \cr
}}

In this description it becomes clear that there may be further surfaces of
singularities when the greatest common divisor of any two of $k,k',k''$ is
not $1$, in analogy with the discussion in \hanur. This will not be
relevant for our purposes and we do not develop the issue further.

After the T-duality, the initial D4 branes become D1 branes located at a
point in the four-fold. When the initial D4 branes are bounded by the grid
of NS, NS$'$ and NS$''$ branes, the T-dual  D1 branes will be located
precisely at the $\IC^4/\Gamma$ singular point. The field theories
introduced in the previous sections correspond to the field theories
appearing in the world-volume of such D1 brane probes. In the following
subsection we will show how the structure of the singularity controls the
spectrum and dynamics of the field theory.

\vskip 1truecm
\subsec{The spectrum of the brane at the singularity}

The field theory on the world-volume of a D brane probes on
singularities can be determined using open string techniques
introduced in \refs{\dm,\dgm}. The particular case of D1 branes on
four-fold quotient singularities $\IC^4/\Gamma$ has been studied in
\mohri, whose results we
review in the present section. We will center on {\it abelian} discrete
groups, and we will also show how the spectra obtained in this
case actually match with those we presented in Section~3. This will
confirm the spectra we had proposed, and also illustrate the usefulness of
the T-duality between brane boxes and branes at singularities.

Let us briefly review the field theory on the world-volume of $N$ D1
branes on $\IC^4$. This is the dimensional reduction of
ten-dimensional $\NN=1$ $U(N)$ super Yang-Mills to two dimensions. The
R-symmetry
group is $SO(8)_R$, which can be understood as the symmetry group of the
transversal dimensions. The theory contains a set of gauge bosons
$A_{\mu}$, singlets under the R-symmetry. The matter
fields are eight real scalar fields $\phi$ transforming in the ${\bf 8_v}$
of $SO(8)$, eight
left-handed fermions $\psi_-$ transforming in the ${\bf 8_s}$, and eight
right-handed fermions $\psi_+$, which transform in the ${\bf 8_c}$. All
fields are in the adjoint representation of the gauge group.

The field theory of D1 branes on $\IC^4/\Gamma$ can be obtained as a
projection of the theory just described onto $\Gamma$ invariant states.
The group $\Gamma$ acts on the R-symmetry of the theory, since this
symmetry associated to the transversal coordinates, which are modded out
to form the quotient. Since $\Gamma$ is a subgroup of $SU(4)$, an unbroken
$U(1)_R$ R-symmetry will remain in the quotient theory, due to the
decomposition $SO(8)_R \supset SU(4) \times U(1)_R$. This action of
$\Gamma$ on $\IC^4$ is specified by a four-dimensional faithful
representation, which we will denote by ${\bf 4}$. Since all irreducible
representations of $\Gamma$ are one-dimensional (recall we only consider
the abelian case), the representation
${\bf 4}$ is reducible, and has a decomposition
\eqn\fourone{{\bf 4} = \RR_{A_1} \oplus \RR_{A_2} \oplus
\RR_{A_3}\oplus \RR_{A_4}}
with the restriction that the tensor product of all
representations is the trivial representation. This is required so that
${\bf 4}$ defines an action in $SU(4)$, rather than in $U(4)$. It also can
be expressed as $A_4=-A_1-A_2-A_3$, in notation of \hanur.

For future convenience, let us point out the this choice of ${\bf 4}$ also
determines a six-dimensional representation ${\bf 6}$, which is obtained
by taking the antisymmetric part of ${\bf 4}\otimes {\bf 4}$
\eqn\sixone{{\bf 6} = \RR_{A_1\oplus A_2} \oplus \RR_{A_1\oplus A_3}
\oplus \RR_{A_2\oplus A_3} \oplus \RR_{-A_1-A_2} \oplus
\RR_{-A_1-A_3} \oplus \RR_{-A_2-A_3}}
where we have introduced the notation $\RR_{A_i}\otimes
\RR_{A_j}=\RR_{A_i\oplus A_j}$ for the tensor product of representations.
The action of $\Gamma$ must also be embedded in the Chan-Paton factors of
the D1 branes. This is specified by giving a representation of $\Gamma$
which we will denote by $\RR_{C.P.}$. The configuration of the D1 branes
is essentially described by such representation.

The simplest physical situation is that of $n$ D1 branes in the quotient
manifold. This can be equivalently described in the cover $\IC^4$ of the
orbifold space as  a set of $n|\Gamma|$ D1-branes (with $|\Gamma|$ being
the order of the discrete group) with Chan-Paton factors in $n$ copies of
the adjoint representation $\RR_{\Gamma}$,

\eqn\chanpaton{\RR_{C.P.} = n \RR_{\Gamma} = n \oplus_I \RR_I.}

The fields in the resulting $(0,2)$ field theory on the D1 branes will be
those invariant under the simultaneous action on R-symmetry and gauge
quantum numbers. Following the rules in \refs{\dm\dgm\vafa\mohri} we can
obtain the
spectrum. The gauge group is $U(n)^{|\Gamma|}$$=\prod_I U(n)_I$, so there
is one gauge factor associated to each irreducible representation of
$\Gamma$.

Since the scalar fields transform in the ${\bf 8_v}$  of $SO(8)_R$, which
decomposes as ${\bf 4}$$+ {\bf \ov 4}$ of the SU(4), to compute the
scalar spectrum in the orbifold theory we have to compute the tensor
products of the representation ${\bf 4}$ \fourone\ with each irreducible
representation $\RR_{I}$,

\eqn\eqchirals{{\bf 4} \otimes \RR_I = \RR_{I\oplus A_1} \oplus
\RR_{I\oplus A_2} \oplus \RR_{I\oplus A_3} \oplus \RR_{I-A_1-A_2-A_3}}

This implies that there are four kinds of complex scalar fields. The
first, associated to
the first complex plane in $\IC^4$, transform as the $(\fund,\antifund)$
of $U(n)_I\times U(n)_{I\oplus A_1}$; we denote these fields as
$\phi_{I,I\oplus A_1}$. The second corresponds to fields denoted
$\phi_{I,I\oplus A_2}$, transforming as the $(\fund,\antifund)$ of
$U(n)_{I}\times U(n)_{I\oplus A_2}$. There are fields $\phi_{I,I\oplus
A_3}$, which transform in the $(\fund,\antifund)$ of $U(n)_I\times
U(n)_{I\oplus A_3}$. Finally, there are fields $\phi_{I,I-A_1-A_2-A_3}$
transforming in the $(\fund,\antifund)$ of $U(n)_I\times
U(n)_{I-A_1-A_2-A_3}$.

The right-handed fermions $\psi_+$ transform in the ${\bf 8_c}$ of
$SO(8)_R$, which also decomposes as ${\bf 4}$$+{\bf\ov 4}$ under $SU(4)$.
To obtain the invariant fields in the orbifold theory one has to use again
the products of the ${\bf 4}$ with the representations $\RR_I$,
\eqchirals.
Thus one obtains four kinds of such fermions fields, denoted
$\psi_{I,I\oplus A_i}$, for $i=1,\ldots,4$, which transform in the
$(\fund,\antifund)$ of $U(n)_I\times U(n)_{I\oplus A_i}$. The scalars
$\phi_{I,I\oplus A_i}$ and the right-handed fermions $\psi_{I,I\oplus
A_i}$ form together $(0,2)$ chiral multiplets, denoted $\Phi^i_{\;I}$ in
the following.

The left-handed fermions $\psi_-$ transform in the ${\bf 8_s}$ of
$SO(8)_R$, whose decomposition under $SU(4)$ is ${\bf 8_s}$$= {\bf 6}$$+
{\bf 1}$$+{\bf 1}$. The two fermions transforming trivially under the
$SU(4)$ will give rise to one left-handed fermion $\chi_I$ (and its
conjugate)
transforming in the adjoint of the gauge factor $U(n)_I$. These fermions,
together with the gauge bosons, form $(0,2)$ vector multiplets. On the
other hand, we have to compute the products of the representation ${\bf
6}$ \sixone\ with the representations $\RR_I$, in order to get the
remaining left-handed fermions. This yields

\eqn\eqfermis{\eqalign{ {\bf 6}\otimes \RR_{I} =  &
\RR_{I\oplus A_1\oplus A_2} \oplus \RR_{I\oplus A_1\oplus A_3}   \oplus
\RR_{I\oplus A_2\oplus A_3} \cr
& \oplus \RR_{I-A_1-A_2}   \oplus \RR_{I-A_1-A_3} \oplus
\RR_{I-A_2-A_3}.\cr}}
Thus we obtain left-handed fermions $\psi_{I,I\oplus A_i\oplus A_j}$, for
$a,b=1,\ldots,4$, $i\neq j$, transforming in the $(\fund,\antifund)$ of
$U(n)_I\times U(n)_{I\oplus A_i\oplus A_j} $. These will correspond to
three Fermi multiplets and their conjugates. We will denote them by
$\Lambda^{ij}_{\;I}$. Again, which fields are considered the Fermi
multiplets, and which their conjugates, depends on the choice of a
`special'
chiral multiplet, to appear in the functions $E$ of the Fermi multiplets.
Following the convention in \mohri, we take $\Phi^4_I$ as these special
fields. This implies that the fields $\Lambda^{i4}_{\; I}$ are the Fermi
multiplets, and that $\Lambda^{ij}_{\;I}$, $i,j\neq 4$ are the conjugates.

The interactions are also given by the projection of the interactions in
the non-orbifolded theory onto $\Gamma$ invariant terms. The outcome can be
recast as a definition of functions $E$ and $J$ for the Fermi multiplets.
The functions $E$ are

\eqn\efunctthree{\matrix{
\Lambda^{14}_{\;I} \;& \to \;\; E^1_{\;I} & =\, \Phi^4_{\;I}
\Phi^1_{\;I-A_1-A_2-A_3} \, -\, \Phi^1_{\;I} \Phi^4_{\;I+A_1}\cr
\Lambda^{24}_{\;I}\; & \to \;\; E^2_{\;I} & =\, \Phi^4_{\;I}
\Phi^2_{\;I-A_1-A_2-A_3} \, -\, \Phi^2_{\;I} \Phi^4_{\;I+A_2}\cr
\Lambda^{34}_{\;I}\; & \to \;\; E^3_{\;I} & = \, \Phi^4_{\;I}
\Phi^3_{\;I-A_1-A_2-A_3} \, -\, \Phi^3_{\;I} \Phi^4_{\;I+A_3}.\cr
}}
And the functions $J$ are

\eqn\jfunctthree{\matrix{
\Lambda^{14}_{\;I} \; & \to \;\; J^{1}_{\;I} & =\,
\Phi^2_{\;I-A_2-A_3} \Phi^3_{\;I-A_3} \, -\, \Phi^3_{\;I-A_2-A_3}
\Phi^2_{\;I-A_2}\cr
\Lambda^{24}_{\;I} \; & \to \;\; J^{2}_{\;I} & =\,
\Phi^3_{\;I-A_1-A_3} \Phi^1_{\;I-A_1} \, -\, \Phi^1_{\;I-A_1-A_3}
\Phi^3_{\;I-A_3}\cr
\Lambda^{34}_{\;I} \; & \to \;\; J^{3}_{\;I} & =\,
\Phi^1_{\;I-A_1-A_2} \Phi^2_{\;I-A_2} \, -\, \Phi^2_{\;I-A_1-A_2}
\Phi^1_{\;I-A_1}.\cr
}}

Observe that the condition $\sum_{i,I} J^i_{\;I} E^i_{\;I}=0$ is verified.
The
interaction terms in components are easily obtained using the equations in
Section~2.

\vskip 1truecm
\subsec{Comparison with the Brane Box Models}

In this section we analyze  the particular case of $\Gamma=\IZ_k\times
\IZ_{k'}\times \IZ_{k''}$, with the generators $\theta$, $\omega$, $\eta$
of $\IZ_k$, $\IZ_{k'}$, $\IZ_{k''}$ acting on $\IC^4$ as in \transf.
In this case the group has $|\Gamma|=k\cdot k'\cdot k''$ irreducible
representations, $\RR_{a,b,c}$, with $a,b,c$ integers defined modulo
$k,k',k''$, respectively. The representation $\RR_{a,b,c}$ associates to
the group element $\theta^l \omega^m \eta^n$ the phase factor
$\exp\,[\,2\pi i ({al \over k} +{bm \over k'} +{cn \over k''})\, ]$.
Clearly,
the tensor product of representations is given by $\RR_{a,b,c}\otimes
\RR_{a',b',c'}$$=\RR_{a+a',b+b',c+c'}$, {\it i.e.} separated addition in
the indices.

With this definition, the action of $\Gamma$ on $\IC^4$ corresponds to the
choice

\eqn\fourtwo{{\bf 4} = \RR_{1,0,0}\oplus \RR_{0,1,0} \oplus \RR_{0,0,1}
\oplus \RR_{-1,-1,-1}.}

From it we can also obtain
\eqn\sixtwo{{\bf 6} = \RR_{1,1,0} \oplus \RR_{1,0,1} \oplus
\RR_{0,1,1} \oplus \RR_{-1,-1,0} \oplus \RR_{-1,0,-1} \oplus
\RR_{0,-1,-1}.}

Using the rules we have described above we can easily find out the
spectrum of the theory. The gauge group is $U(n)^{kk'k''}=\prod_{a,b,c}
U(n)_{a,b,c}$, where each factor is associated to an
irreducible representation. Let us now look at the four types of $(0,2)$
chiral multiplets $\Phi_{I,I\oplus A_i}$, $i=1,\ldots,4$. The fields
associated to the first complex plane, $\Phi^1_{a,b,c}$, transform in the
$(\fund,\antifund)$
of $U(n)_{a,b,c}\times U(n)_{a+1,b,c}$. The fields associated to $A_2$,
$\Phi^2_{a,b,c}$, transform in the $(\fund,\antifund)$ of
$U(n)_{a,b,c}\times U(n)_{a,b+1,c}$. The fields $\Phi^3_{a,b,c}$ transform
in the $(\fund,\antifund)$ of $U(n)_{a,b,c}\times U(n)_{a,b,c+1}$.
Finally, the fields $\Phi^4_{a,b,c}$ transform in the $(\fund,\antifund)$
of $U(n)_{a,b,c}\times U(n)_{a-1,b-1,c-1}$.

The Fermi multiplets are obtained using the representation ${\bf 6}$
above. The field $\Lambda^{14}_{a,b,c}$ transforms in the
$(\fund,\antifund)$ of $U(n)_{a,b,c}\times U(n)_{a,b-1,c-1}$; the fields
$\Lambda^{24}_{a,bc}$ transform in the $(\fund,\antifund)$ of
$U(n)_{a,b,c}\times U(n)_{a-1,b,c-1}$; the fields $\Lambda^{34}_{a,b,c}$
transform in the $(\fund,\antifund)$ of $U(n)_{a,b,c}\times
U(n)_{a-1,b-1,c}$.

We have listed the spectrum of chiral multiplets and Fermi multiplets in
Table~7.

\midinsert{
{\sevenrm
$$
\vbox{\offinterlineskip
\def\strut{\vrule height 3.25ex  width 0pt depth 2ex}
\def\hline{\noalign{\hrule}}
\halign{\vrule#\hfil\strut&\quad#\hfil\quad\vrule&&\quad\enspace
\hfil#\hfil\quad\vrule\cr
\noalign{\hrule}
&$\hfil {\bf Field}$& ${\bf Representation}$ & {\bf Brane box}\cr
\noalign{\hrule}
\hline
&$\Phi^1_{a,b,c}$& $(\fund_{a,b,c},\antifund_{a+1,b,c}$ & $H_{a,b,c}$ \cr
\hline
&$\Phi^2_{a,b,c}$& $(\fund_{a,b,c},\antifund_{a,b+1,c})$ & $V_{a,b,c}$\cr
\hline
&$\Phi^3_{a,b,c}$& $(\fund_{a,b,c},\antifund_{a,b,c+1})$ & $N_{a,b,c}$\cr
\hline
&$\Phi^4_{a,b,c}$& $(\fund_{a,b,c},\antifund_{a-1,b-1,c-1})$ &
$D_{a,b,c}$\cr
\hline
\hline
&$\Lambda^{14}_{a,b,c}$& $(\fund_{a,b,c},\antifund_{a,b-1,c-1})$
& $\Lambda^{({\bar 2})}_{abc}$\cr
\hline
&$\Lambda^{24}_{a,b,c}$& $(\fund_{a,b,c},\antifund_{a-1,b,c-1})$
& $\Lambda^{({\bar 1})}_{a,b,c}$\cr
\hline
&$\Lambda^{34}_{a,b,c}$& $(\fund_{a,b,c},\antifund_{a-1,b-1,c})$
& $\Lambda^{(3)}_{a,b,c}$\cr
\hline
}}
$$ }}
{\ninerm
{Table 7: The matter content of the theory of D1 branes at
singularities. In the last column we compare these fields with the
spectrum obtained in a $k\times k'\times k''$ brane box model of
Section~3.}}
\endinsert

Notice that the spectrum we have obtained corresponds precisely to the one
we proposed for a $k\times k'\times k''$ box model with trivial
identifications of sides. The chiral multiplets $\Phi^i_{a,b,c}$
correspond to the
fields $H$, $V$, $N$, $D$ of Section~3. The choice of $\Phi^4$ as
special field corresponds to our convention of section~3.3, of choosing
the $D$ fields as special. The Fermi multiplets $\Lambda^{14}$,
$\Lambda^{24}$, $\Lambda^{34}$ in the singularity correspond to the Fermi
fields $\Lambda^{({\bar 2})}$, $\Lambda^{({\bar 1})}$, $\Lambda^{(3)}$,
respectively. This comparison is shown in the last column of Table~7.
It is also a simple matter to particularize expressions
\efunctthree\ and \jfunctthree\ to this case and reproduce the interactions
proposed in Section~3.3

The matching of the two field theories was actually expected, since we had
already
argued that the configurations are related by T-duality. The detailed
correspondence of the spectra and interactions in both constructions
is a nice argument supporting both the T-duality, and the spectrum proposed
in Section~3 for the brane box configuration.

The T-duality proposal can be generalized in several directions. For
instance, there is no difficulty in extending it to models with
non-trivial identifications of faces of the unit cell, following \hanur.
Also, by considering fractional branes at the singularity one finds the
T-duals of brane box models with different number of D4 branes in each
box.

Rather than commenting on these exercises, 
we turn to briefly mention a few facts about the phase structure (in the
sense of \witten) of these
field theories as linear sigma models. Even though the field theories are
relatively complicated, this analysis has been carried
out for $(2,2)$ theories \dgm, and $(0,2)$
theories \mohri\foot{The theories in
these references actually correspond to the case where all the gauge
factors are $U(1)$'s.}. The outcome is that only the geometric phases of
the linear sigma model are realized, and so no CY/LG correspondence holds
in this case. However, flop transitions between different geometric
phases \agm\ are certainly possible \flops. Since the field theories in
this section are the same as those arising from brane box models (via the
T-duality map), the results mentioned above also apply in the brane box  
context.

In the following Section we illustrate another interesting application of 
the T-duality map. We turn to the study of
the models with enhanced supersymmetries introduced in Section~4.

\subsec{Enhanced Supersymmetries}

When the discrete group $\Gamma$ that defines the singularity is not a
generic subgroup of $SU(4)$, so that a larger R-symmetry is left unbroken,
we can expect an enhancement of supersymmetry. The simplest such examples,
yielding  non-chiral supersymmetry enhancement appear when $\Gamma\subset
SU(3)$. The  R-symmetry of the theory is $U(1)\times U(1)$, and the
theories obtained are $(2,2)$ in two dimensions. These models are clearly
T-dual of configurations of D3 branes at Calabi-Yau three-fold
singularities $\IC^3/\Gamma$, which yield
four-dimensional $\NN=1$ theories. Analogously, when $\Gamma\subset
SU(2)$, the $(4,4)$ theories obtained are related of six-dimensional
$\NN=1$ theories, which can be realized in the world-volume of D5 branes
at ALE singularities $\IC^2/\Gamma$.

The relation with the brane box model is clear. When the singularity has a
restricted holonomy group, $SU(3)$ (resp. $SU(2)$), the brane box
configuration reproducing the $(2,2)$ (resp. $(4,4)$) field theory has
only two (resp. one) kinds of NS fivebranes. Thus the brane box
configurations also admit an interpretation as reduction of higher
dimensional brane models.

\medskip

More interesting is the chiral enhancement of supersymmetry. As mentioned
in \mohri, one can obtain $(0,4)$ theories in the D1 brane picture when
the four-fold singularity admits some additional covariantly constant
spinors. This can be translated in the condition $\Gamma\subset
SU(2)\times SU(2)$, which leaves a $SU(2)\times SU(2)$ unbroken
R-symmetry. This restricted holonomy implies the $\Gamma$-invariance of
the forms $dz_1\wedge dz_2$ and $dz_3\wedge dz_4$ in $\IC^4$. This
condition restricts the possible actions of $\Gamma$ on $\IC^4$, in such a
way that it ensures the existence of one Fermi multiplet in
the adjoint. This fermions are required in a $(0,4)$ theory since they
become part of the $(0,4)$ vector multiplet. Also, one can check that the
chiral multiplets arrange in conjugate pairs to form $(0,4)$ chiral
multiplets. The interactions also respect the higher supersymmetry.

These and some of the following special holonomies have also appeared in
the study of two-dimensional compactifications of string theory and
M-theory (see for example \dasmuk).
A particular family of such singularities is mentioned in \mohri. The
group is $\IZ_n$, with the generator $\theta$ acting on $\IC^4$ as
\eqn\action{\theta:\; (z_1,z_2,z_3,z_4)\;\to\; (e^{2\pi i/n}
z_1,e^{-2\pi i/n}z_2,e^{2\pi i a/n}z_3, e^{-2\pi ia/n}z_4)}
It is a simple matter to obtain the spectrum and interactions in this
example. One can also easily find brane box configurations yielding the
same field theory. In fact, the example shown in Figure~8 correspond to the
$\IZ_4$ $n=4,a=2$ model in the family above.

Another interesting family of such theories is obtained as the product of
two ALE singularities living in the $(z_1,z_2)$ and $(z_3,z_4)$ spaces,
respectively.

Let us stress that for {\it any} singularity yielding a $(0,4)$ theory we
can find brane box configurations reproducing the same field theory
\foot{As in \hanur, there are in general several different brane box
models corresponding to the same field theory. They correspond to
T-dualizing the same singularity along different sets of directions.}.
The construction is
systematic and closely analogous to that discussed in \hanur\ for
three-fold singularities, so we will not discuss it here.

Observe that the basic requirement in order to have $(0,4)$ supersymmetry
can be rephrased as the existence of Fermi multiplets in the adjoint
representation. This agrees with our comments from the brane box point of
view in section~4.2.

\medskip

Concerning the field theories with $(0,6)$ supersymmetry, they are also
obtained using singularities with restricted holonomy. In this case the
condition amounts to ensuring the existence of two $(0,2)$ Fermi
multiplets in the adjoint representation (and no adjoint chiral
multiplets). A complete classification of such singularities is possible,
the only such models are $\IZ_n$ orbifolds  with the generator $\theta$
acting as
\eqn\actiontwo{\theta:\; (z_1,z_2,z_3,z_4)\;\to\; (e^{2\pi i/n}
z_1,e^{-2\pi i/n}z_2,e^{2\pi i/n}z_3, e^{-2\pi i/n}z_4).}
The unbroken R-symmetry is $SU(4)_R\approx SO(6)_R$. Again it is a simple
exercise to
obtain the spectrum and interactions of the theory, and find brane box
models yielding the same field theory. As an example, let us mention that
the four-box model shown in Figure~9 gives precisely the field theory of
the $\IZ_4$ example in the family just mentioned. The general $\IZ_n$ case
is obtained by considering an analogous brane box model with $n$ boxes. As
indicated in section~4.2, the spectrum in the general case is given in
Table~6.

\medskip

Finally, let us comment on the only singularity yielding a $(0,8)$
supersymmetric theory. It is the $\IZ_2$ orbifold whose generator $\theta$
acts as inversion of the four coordinates in $\IC^4$. The unbroken
R-symmetry is $SO(8)_R$. As for the spectrum, the gauge group is $U(n)^2$,
all the Fermi multiplets transform in the adjoint, and all chiral
multiplets transform in bi-fundamental representations. The model is
clearly the same one we discussed in section~4.2, and corresponds to the
brane box configuration shown in Figure~10.

\medskip

The main conclusion of this discussion is that enhanced supersymmetry is
associated to a
restricted holonomy in the four-fold singularity. The T-duality map
discussed in previous subsection then seems to suggest that the  T-dual
brane configurations present enhanced supersymmetry due to properties of
their `generalized holonomy group' in the sense of \angles. In the cases
of non-chiral supersymmetry enhancement this is obvious, since essentially
the correspond to removing one kind of brane. In the more interesting case
of chiral supersymmetries, on the other hand, the
supersymmetry enhancement is not so obvious in the brane box
configuration and only follows upon the detailed computation of spectrum
and interactions. It would be nice to achieve a better understanding of
such brane box models.

\vskip 2truecm


\newsec{Concluding Remarks}

In this paper we have introduced brane box configurations giving chiral
field theories in two dimensions. An interesting feature is that are a
natural generalization of brane models yielding chiral field theories in
six and four dimensions. The connection between the chiral theories in the
different dimensions is dimensional reduction followed by an appropriate
chiral projection, performed by the introduction of a new kind of NS
fivebrane.

We have provided the rules to determine the field theory spectrum from the
brane box data. Since the theories have only two supersymmetries, the
interactions have a relatively complicated structure. However, we have
shown how they can be easily encoded in a diagrammatic representation in
the brane box configuration.

We have also discussed the conditions under which the models present
enhanced supersymmetry. We have stressed that it is possible to construct
brane box models with enhanced $(0,4)$, $(0,6)$ and $(0,8)$ supersymmetry,
and shown several examples of such configurations.

In the case where the direction 246 in the brane configuration are
compact, we have related the corresponding models to theories arising
from D1 brane probes at $\IC^4/\Gamma$ singularities, with $\Gamma$ an
abelian subgroup of $SU(4)$. The relation is based on T-duality along the
compact directions. This result has been useful, since it provides a
simple rederivation of the rules to compute the spectrum and interactions
we had proposed in the brane box picture. It also provides a nice
geometrical interpretation for the enhancement of supersymmetry (both
chiral and non-chiral) in terms of the holonomy group of the four-fold.
Much could be said about the relation of the brane boxes and the
singularity theories. However, since this discussion is essentially
identical to that in \hanur, we have not repeated the details. Let us
stress however that there is no difficulty in repeating the exercise in
the four-fold case.

There are several interesting results concerning the phase structure of
these field theories, when understood as linear sigma models
\refs{\dgm,\mohri}. There the field theories appeared from D-branes at
singularities. An interesting application of the T-duality map is that it
ensures these are the same theories one obtains from brane box
configurations. This allows us to directly borrow the corresponding
results and learn that only the geometric phases are accesible to  
the field theories. However, flop phase transition are still possible, and
it would be interesting to give them a direct intepretation in the brane  
box picture.

Another interesting point is that this relation between brane boxes and
branes at singularities is expected to shed light on some issues which
are not completely clear in the brane box construction, such as the
restrictions in the number of D4 branes that one is allowed to put in the
boxes. Such restrictions are required to ensure cancellation of gauge
anomalies, and can be presumably studied in the singularity picture in the
context of perturbative string theory. We leave this as an open question.

Observe that the singularity picture provides the construction of a larger
family of $(0,2)$ theories, by considering non-abelian subgroups of
$SU(4)$. The rules to compute the spectrum are a simple generalization of
those we mentioned in the abelian case.

Finally let us mention an amusing relation between the $(0,2)$ theories we
have studied and the dimensional reduction of the four-dimensional
non-supersymmetric orbifold theories introduced in \ks. As we
have mentioned, the theory of D1 branes on $\IR^8$ has $SO(8)_R$
R-symmetry. The scalar fields $\phi$ transform in the ${\bf 8_v}$
representation, the right-handed spinors $\psi_+$ transform in the ${\bf
8_c}$, and the left-handed spinors transform in the ${\bf 8_s}$.

There are three inequivalent embeddings of $SU(4)$ in $SO(8)$. They differ
in which of the three representations `${\bf 8}$' decomposes as ${\bf
6}$$+{\bf 1}$$+{\bf 1}$. Since the three representations are related by
$SO(8)$ triality, so are the three embeddings. Taking this into account,
given a (possibly non-abelian) discrete group $\Gamma \subset SU(4)$, and
four- and six- dimensional representations ${\bf 4}$ and ${\bf 6}$, there
are three possible orbifold theories we can define, using the three
embeddings of $SU(4)$ in $SO(8)$.

We have already met one of these, where ${\bf 8}_s$ decomposes as
${\bf 6}$$+{\bf 1}$$+{\bf 1}$, and both ${\bf 8_v}$ and ${\bf 8_c}$
decompose as ${\bf 4}$$+{\bf{\bar 4}}$. The orbifold theory is a $(0,2)$
chiral supersymmetric theory. It contains scalars $\phi_{I,I\oplus A_i}$,
right-handed fermions $\psi^+_{I,I\oplus A_i}$, and left-handed fermions
$\psi^-_{I,I\oplus A_i\oplus A_j}$ (in notation of Section~5). The
right-handed spinors and scalars
are related by the right-handed supersymmetries. The conmutant of the
$SU(4)$ is the $U(1)$ R-symmetry of the theory.

Another embedding, in which ${\bf 8}_c$ decomposes as ${\bf 6}$$+{\bf 1}$
$+{\bf 1}$, and ${\bf 8}_v$, ${\bf 8}_s$ decompose as ${\bf 4}$$+{\bf
{\bar 4}}$, is clearly the parity transformed of the precedent. The
scalars
$\phi_{I,I\oplus A_i}$ and the left-handed fermions $\psi^-_{I,I\oplus
A_i}$ are related by $(2,0)$ supersymmetry. There are also right-handed
fermions $\psi^+_{I,I\oplus A_i\oplus A_j}$.

Finally there is a third choice, in which ${\bf 8}_v$ decomposes as
${\bf 6}$$+{\bf 1}$$+{\bf 1}$, and ${\bf 8}_s$, ${\bf 8}_c$ decompose as
${\bf 4}$$+{\bf{\bar 4}}$. The field theory contains real scalars
$\phi_{I,I\oplus A_i\oplus A_j}$, complex scalars $\phi_{I,I}$, and
vector-like fermions $\psi^{\pm}_{I,I\oplus A_i}$. These theories are the
dimensional reduction of the four-dimensional non-supersymmetric theories
introduced in \ks. Now the $U(1)$ which is unbroken arises
in the dimensional reduction. Observe the spectrum is similar to that of
the $(0,2)$ or $(2,0)$ theories by an exchange of the scalars with one
kind of fermions. This is somewhat reminiscent of the twisting of
supersymmetric theories where one mixes the Lorentz group and the internal 
symmetries of the theory, even though we have not been able to make this
more precise. It is tantalizing to imagine that this amusing group
theoretical fact may have some deeper meaning.

\medskip

\centerline{\bf Acknowledgements}

It is a pleasure to thank A.~Hanany and R.~Blumenhagen for useful comments.
H. G-C. thanks to E.~Witten for hospitality in the Institute for Advanced
Study. A.~U. is indebted to M.~Gonz\'alez for her support. The work H. G-C.
is supported by a Postdoctoral CONACyT fellowship (Mexico) under the
program {\it Programa de Posdoctorantes: Estancias Posdoctorales
en el Extranjero para Graduados en Instituciones Nacionales 1996-1997}.
The work of A.~U. is supported by a fellowship from the Ram\'on Areces
Foundation (Spain), and partially by the CICYT (Spain), under grant
AEN97-1678.

\vskip 2truecm


\appendix{I}{Parameters in the Model: Gauge Couplings, Theta Parameters,
Fayet-Iliopoulos Parameters}

Gauge couplings, theta angles and Fayet-Iliopoulos (FI) parameters can be
characterized in analogy with the four-dimensional brane box
configurations \refs{\zaffa,\math}.

\noindent
{\it Gauge couplings}

The gauge couplings of the
various gauge groups are given in terms of the positions of the NS branes
along $x_6$
direction, the positions of the NS' branes along the $x_4$ direction and the
positions of the NS'' branes along the $x^2$ direction. There are $k$ possible
positions $x^i_6$, $k'$ positions $x^j_4$ and $k''$ $x^k_2$. These directions
are divided into $k$ intervals of lengths $I_a=x^a_6 - x^{a-1}_6$ for
$x_6$,
into $k'$ intervals of lengths $J_b=x^b_4 - x^{b-1}_4$ for $x_4$ and
into $k''$ intervals of lengths $K_c=x^c_2 - x^{c-1}_2$ for $x_2$. The intervals
$I_a,J_b$ and $K_c$ satisfy the relations $\sum_a I_a = R_6$,
$\sum_b J_b = R_4$ and $\sum_c K_c = R_2$. Thus the gauge coupling $g_{a,b,c}$
associated to the $U(n_{a,b,c})$ is given by

\eqn\coupling{{1 \over g^2_{a,b,c}} = { I_a J_b K_c\over g_s \ell_s}.}

\medskip

\noindent
{\it Theta angles}

The interaction which gives rise to the theta angle of the
gauge theories in two dimensions can be seen to arise from the
following interaction in the world-volume of the M-theory fivebrane

\eqn\interaction{ \int_{W_6} C^{(3)} \wedge  dB^{(2)}.}
It is well known that the fivebrane of M-theory wrapped on ${\bf S}^1$
is precisely the $D4$-brane of type IIA superstring theory. Thus the fivebrane
will extend in the directions $x_0,x_1,x_2,x_4,x_6,x_{10}$ with $x_{10}$
parametrizes ${\bf S}^1$. Interaction \interaction\ turns out to be

\eqn\ident{ \int_{(01246) \times {\bf S}^1} C^{(3)}  \wedge
dB^{(2)}, }
with $(01)$ the world-sheet coordinates where is defined the two-dimensional
gauge theory. Under compactification on the circle $B^{(2)}$ field becomes
the one-form $A$ on the world-volume of the D4 branes. Eq. \ident\
yields
\eqn\more{ \int_{(01246)} C^{(3)}  \wedge dA, }
Under the further dimensional reduction of the D4 brane on 246 we obtain
the two-dimensional field theory theta terms
\eqn\final{ \theta \int_{(01)} F , }
where $\theta = \int_{(246)} C^{(3)}$.

There is another equivalent way to obtain the interaction \more, from the
Chern-Simons contribution to the D-brane Lagrangian \refs{\douglas}

\eqn\cs{ \int_{W_{p+1}} C \wedge {\rm Tr} e^{\cal F},}
where $C$ is the sum over all RR fields of the relevant string theory
and ${\cal F}$ is the field strength of the gauge fields on $W_{p+1}$. For
D4 brane of type IIA string theory Chern-Simons Lagrangian looks like

\eqn\csiia{ \int_{W_5} \bigg( C^{(1)} {\rm Tr}({\cal F}\wedge
{\cal F})  +  C^{(3)} Tr {\cal F}\bigg),}
with $C^{(1)}$ and $C^{(3)}$ the IIA one- and three-forms. The last term
reproduces Eq.\more.

The above analysis gives the correct answer only for the case of D4 branes
wrapped on a ${\bf T}^3$, which gives two-dimensional gauge theories with
sixteen supercharges. In more general cases, the theta angle receives
further contributions, which we analyze in the following.

Let $R_{a,b,c}$ denote the volume of a $246$ box which is bounded by
NS, NS$'$ and NS$''$ branes.
We define $Q_a$ to be the face contained in the $a^{th}$ NS brane, $Q_{b}$
the face contained in the $b^{th}$ NS$'$ brane, and $Q_{c}$ the face in
the $c^{th}$ NS$''$ brane.
Let us consider the two-form fields $B^{(2)}$, $B'^{(2)}$ and $B''^{(2)}$
on the world-volume of NS, NS$'$ and NS$''$ branes.  The theta angle for
the group associated to the box $(a,b,c)$ depends on the surface integrals
along the boundary of $R_{a,b,c}$.

Let us first consider models with $(4,4)$ supersymmetry in two
dimensions. We can construct such models by considering $k'=k''=0$, and
246 compact. Each factor in the group arises from a box with the topology
of a two-torus (along 2,4) times an interval (bounded by the positions of
NS branes along 6). In such situation, the definition
$\theta=\int_{R_{a,b,c}} C^{(3)}$ is not invariant under the gauge
transformation of the three-form $C^{(3)}\to C^{(3)}+d\lambda$, with
$\lambda$ a two-form, since $\theta$ would pick a boundary term. This
problem is solved by refining the definition as follows

\eqn\modif{ \theta_{a} = \int_{R_{a,b,c}} C^{(3)} + \int_{Q_{a}}(
B^{(2)}_a - B^{(2)}_{a-1}).}
and allowing $B^{(2)}$ to transform as $B^{(2)}\to B^{(2)}+\lambda$.

In the general case in which the box $R_{a,b,c}$ is bounded by NS
fivebranes in all three directions 246, the expression for the theta angle
includes all the boundary terms.

\eqn\formula{ \theta_{a,b,c} = \int_{R_{a,b,c}} C^{(3)}
+ \int_{Q_{b}} (B^{(2)}_b -B^{(2)}_{b-1})
+ \int_{Q_{a}} (B'^{(2)}_{a-1} - B'^{(2)}_{a})
+ \int_{Q_{c}} (B''^{(2)}_c -B''^{(2)}_{c-1})}

Now this expression is not invariant under the gauge transformations of
the two-form fields, $B^{(2)}_a\to B^{(2)}_a + d\alpha_a$, etc, (here the
$\alpha$'s are one-forms). Under this transformations, $\theta$ would pick
up terms at the boundaries of the $Q$'s, {\it i.e} intervals where two
kinds of NS fivebranes intersect. This forces a further refinement by
introducing integrals of appropriate one-form fields along those
intervals. Finally, there should exist a contribution from the points of
intersection of the three kinds of NS fivebranes, in order to ensure
invariance under gauge transformations of the one-forms.
We have not written these last contributions in the formula \formula.

\medskip

\noindent
{\it Fayet-Illiopoulos parameters}

In two dimensions, we expect the gauge groups to contain $U(1)$ factors, and
so the
two-dimensional field theory contains Fayet-Iliopoulos parameters.
These parameters  will be associated to the positions of fivebranes in
several directions. There are three contributions
to the FI parameter of a given gauge group,
one for each type of NS fivebrane bounding the corresponding box.
One contribution comes from the position of the NS branes in $x^7$.
For example, consider the case of $k$ arbitrary and $k'=k''= 1$. For each one
of the $k$ boxes we have a U(1)-factor and correspondingly a FI parameter.
This is given by
\eqn\fayet{r_a = (x^7)_{NS_a} - (x^7)_{NS_{a+1}}\equiv \delta(x^7)_{a}.}
These FI parameters fulfill the condition $\sum_a^{k-1} r_a = 0$
required in order to have unbroken supersymmetry. The motivation for
equation \fayet\ is that the Higgs
breaking triggered by the contribution of the FI to the D-term is
precisely that obtained upon moving the NS brane in $x^7$.

This argument can be extended to the general case with arbitrary number
of boxes. In this case FI parameters are given in terms of the
differences of positions of NS branes in $x^7$, NS$'$ branes in
$x^5$, and NS$''$ branes in $x^3$, as follows
\eqn\general{r_{a,b,c}= (\delta x^7)_{a} + (\delta x^5)_{b}
+ (\delta x^3)_{c}.}
They verify the condition $\sum_{a,b,c} r_{a,b,c} = 0$.

\medskip

In general we expect that, as happened with the theta angles, the field
theory parameters will receive contributions associated to the
intersections of NS fivebranes. This was already observed in \math, and
is expected to happen in our case in even a more complicated manner.

\listrefs

\end